\documentclass[aps,prl,groupedaddress,showpacs]{revtex4}

\usepackage{bm}
\usepackage{amssymb}
\usepackage[dvips]{graphicx}

\begin{document}


\title{Direct calculation of interfacial tensions \\  from computer simulation: \\
Results for freely jointed  tangent hard sphere chains}


\author{Luis G. MacDowell}
\affiliation{Departamento de Qu\'{\i}mica F\'{\i}sica, Facultad de Ciencias Qu\'{\i}micas,
Universidad Complutense de Madrid, 28040, Spain.}
\author{Pawe{\l}  Bryk}
\affiliation{Department for the Modeling of Physico--Chemical Processes, Maria Curie--Sklodowska, 20--031
Lublin, Poland.}

\newcommand{\prom}[2]{\left \langle #1 \right \rangle_{#2}}
\newcommand{\du}{\mbox{$\,$} \mathrm{d}}
\newcommand{\Ppar}{p_{\parallel}}
\newcommand{\Pper}{p_{\perp}}
\newcommand{\derpar}[3]{
        \left(\frac{\partial #1}{\partial #2}\right)_{#3}
                      }
\newcommand{\Eq}[1]{Eq.~(\ref{#1})}
\newcommand{\rvec}[2]{\mathbf{r}_{#1}^{#2}}
\newcommand{\qvec}[2]{\mathbf{q}_{#1}^{#2}}
\newcommand{\tvec}[2]{\mathbf{t}_{#1}^{#2}}


\date{\today}

\begin{abstract}
We develop a  methodology for the calculation of surface free energies based on the probability 
distribution of a wandering interface. Using a simple extension of the NpT sampling, we allow the interface
area to randomly probe the available space and evaluate the surface free energy from histogram analysis and the
corresponding average. The method is suitable for studying systems with either continuous or discontinuous
potentials, as it does not require explicit evaluation of the virial. The proposed algorithm is compared
with known results for the surface tension of Lennard--Jones and Square Well fluid, as well as for the
interface tension of a bead--spring polymer model
and good agreement is found. We also calculate interfacial tensions of freely jointed
tangent hard sphere chains on athermal
walls for a wide range of chain lengths and densities. The results are compared with three different
theoretical approaches, Scaled Particle Theory, the Yu and Wu density functional theory and an analytical
approximation based on the latter approach. Whereas SPT only yields qualitative results, the last two
approaches are found to yield very good agreement with simulations.
\end{abstract}


\maketitle

\section{Introduction}

Interfacial phenomena have been a matter of great research interest for centuries \cite{rowlinson82b}.
With the growing importance of nanotechnology, however, this field of research is expected to become even more
relevant. As the surface to volume ratios diminish in miniaturized devices, surface effects become
increasingly important and strongly influence the state of the systems \cite{seemann05}, including new and 
surprising behavior \cite{milchev05,macdowell06}.

Whereas the structure of the interface can be quite complex at the microscopic
level \cite{rowlinson79,cahn58,tarazona84,woodward90,mueller00,bryk04,bryk04b}, from a thermodynamic point of view only
the interfacial tensions are really required to describe the system's behavior in most instances (e.g., Young's equation).
Statistical mechanics provides a link for the calculation of interface tensions in terms of the average
of a mechanical property, namely, the anisotropy of the pressure tensor \cite{irving50}. This route has been
applied  to estimate interface tensions, either theoretically \cite{rowlinson82b}, or by means of
computer simulations \cite{magda85}. Results are known for some prototype systems, including 
the
Lennard--Jones \cite{trokhymchuk99}, Square Well \cite{orea03} and Gay--Berne \cite{martin97} fluids,  
as well as  bead--spring polymer
chains \cite{varnik00,milchev01,duque04}.
Unfortunately, the explicit evaluation of the pressure tensor is already a very subtle matter for simple
fluids \cite{rowlinson82b,varnik00}. The presence of discontinuities in the potential, either inherent in the model
or resulting from truncation in computer simulations can cause difficulties \cite{trokhymchuk99}, and at any
rate need the approximate evaluation of Dirac {\em delta} functions. For molecular fluids the difficulties may
become even more important, since the bonding is often rigid and the resulting contribution of the
constraining forces becomes difficult to estimate \cite{duque04}.

Several methods have been proposed in the literature to overcome the difficulties related to  explicit
evaluation of the virial. A useful methodology which is particularly interesting in the vicinity of the
critical point was proposed long time ago by Binder \cite{binder82}, and has received renewed interest
recently \cite{mueller00,potoff00,errington03}. Although cast in terms of probability distributions,
this method is equivalent to   a full thermodynamic integration from the one phase state to a two phase
state \cite{macdowell03b}.
In that sense, it can  be considered as an
extension of thermodynamic integration employed to calculate interfacial tensions from adsorption 
isotherms \cite{hooper00}.
The need for sampling all states between the homogeneous bulk phase and the phase coexisting states (slab
geometry) makes it fairly time consuming away from the critical point. On the other hand,  
one gets a large amount of information on the intermediate metastable states which can
be exploited  for use in nucleation studies \cite{furukawa82,macdowell04}. 
Alternatively,
one can explicitely simulate two phase states only and attempt to collect averages containing information on
the interface. One such method is based on the capillary Hamiltonian approximation, whereby the spectra of
interfacial capillary waves is analyzed, and the surface tension is obtained from the long wavelength
behavior \cite{mueller96,mueller00,milchev01}. This method may be applied only for the liquid--vapor interface
and presents difficulties related to the explicit location of the
interface boundary, but does not seem to be affected by this arbitrariness at  long
wavelengths \cite{chacon05}.
A third route that allows to calculate interfacial tensions without explicit evaluation of the virial is based on
the perturbation of the two phase system, by increasing or decreasing the interface a finite amount. For
curved interfaces, this method resembles the philosophy of Scaled Particle
Theory \cite{reiss59,henderson83,bryk03},
and was exploited by 
Bresme and Quirke in order to calculate the interfacial tension of colloidal nano particulates \cite{bresme98,bresme99}.
Another
perturbative approach, the Test Area Method of Gloor et al., has been recently applied to the study of 
flat interfaces,
providing reliable estimates for spherical and anisotropic fluids \cite{gloor05,demiguel06}. These methods are special
cases of a more general methodology which can be applied to measure first derivatives of the free
energy \cite{vortler00}, with the  Widom Test particle method as the most widely known
application \cite{widom63}. In the perturbative methods,
the difficulty is the choice of surface increment, and the reliable average of the resulting small energy change, as
is the case in the evaluation of numerical derivatives.
In this work we propose a new method which has features of both the Binder method, in the sense that it relies
on histogram analysis, and the perturbative approaches employed by Bresme and Quirke and Gloor et al., in the sense that
the system is subject to small perturbations of the surface area. The difference is that the perturbations 
add up and the interface is
allowed to wander freely. An analysis of the resulting probability distribution, or, alternatively, 
the corresponding average, yields the required interface tensions.

Another route for calculating the interfacial tension is the density functional theory.
In this approach the grand potential of the system is a functional of the local density,
therefore the interfacial tension is readily accessible once the functional is minimized.
Several density functionals and related self consistent field theories  for polymeric fluids 
have been proposed in the literature \cite{helfand72,Chandler86:1,woodward90,mueller03b}.
One class of functionals is based on the polymer reference interaction site model.
\cite{Chandler86:1,hooper00:1,hooper00,Frischknecht02:1}
Another class is based on the Woodward formalism \cite{woodward90,yu02,kierlik93}.
In this work we use the functional belonging to the latter class and proposed by
Yu and Wu \cite{yu02}. This theory is based on Wertheim's first-order thermodynamic
perturbations theory\cite{wertheim87} (TPT1),
and is one of the most accurate functionals for tangent hard sphere systems.
However, besides the comparison with our new simulation
methodology our choice of the functional serves another purpose.
The Yu and Wu theory is also used as a starting point for
derivation of an analytical approximation for the interface  tension of
tangent hard-sphere chains at hard walls. This simple approximation
will turn out to be in a good agreement with the simulation data.

The content of the paper is organized as follows.
In next section we briefly consider the surface thermodynamics of systems with inhomogeneous density profiles
along one direction perpendicular to the interface. This applies for two cases of interest, free liquid--vapor
interfaces and fluid--substrate interfaces. Having considered this formal aspects, we present the new method
proposed for the calculation of interface tensions in section III. Results for the different
systems studied, from the Lennard--Jones fluid to athermal chains is presented in section IV, where
comparison is made to DFT results and Scaled Particle Theory. Finally, section V presents our conclusions.

\section{Surface Thermodynamics of Slit Pore Geometry}

In this section we will study the thermodynamics of systems having an inhomogeneous
interface profile along a single direction, $z$.  This includes two systems of interest,
among others. One consists of (say) a liquid slab surrounded by vapor (or any other coexisting phase)
in a system with periodic boundary conditions in all three directions. The other is
a fluid in contact with a substrate, with periodic boundary conditions only in the
directions perpendicular to $z$. For the sake of generality, let the container  have orthogonal shape, with sides
 of length $L_x$, $L_y$ and $L_z$ (note that in a slit pore the choice of $L_z$ amounts to an arbitrary
 definition of the dividing surface).
Work can be done on the system by changing the shape in either of two ways.
Firstly, one can perform displacements of the $xy$ plane along  the perpendicular direction. 
Let $\Pper$ be the external pressure exerted on this moving plane.
The work done by an infinitesimal increment of $L_z$ is then given by $\du w=-L_xL_y \Pper \du L_z$.
Alternatively, one can perform perpendicular displacements of the $yz$ and $zx$ planes, thereby producing
an infinitesimal deformation  of the $xy$ plane. The resulting amount of work is then
$\du w= - L_z \Ppar \du (L_xL_y)$,
where $\Ppar$ is the average external pressure exerted on the $yz$ or $zx$ planes:
\begin{equation}
  \Ppar = \frac{1}{L_z}\int \frac{1}{2}\left( p_{xx}(z) + p_{yy}(z) \right ) \du z  
\end{equation} 
Accordingly, we may write for the infinitesimal changes in energy:
\begin{equation}
 \du U =  T\du S - A\Pper \du L_z - L_z \Ppar \du A + \mu \du N
\end{equation} 
where $A=L_xL_y$.
Introducing a grand potential by means of the Legendre transform
$\Omega = U - TS - \mu N$ yields:
\begin{equation}\label{domega}
\du \Omega = - S \du T - N \du\mu - A\Pper \du L_z - L_z \Ppar \du A 
\end{equation} 

In practice, we will be concerned with deformations which preserve the
volume, hence $A\du L_z + L_z \du A = 0$. It then follows that
\begin{equation}
 \derpar{\Omega}{A}{\mu VT} = - L_z(\Ppar - \Pper)
\end{equation} 
In the limit of large $L_z$, where the two interfaces do not interact any longer,
the above derivative is identified with twice the interface tension, $\gamma_{\infty}$. 
For the more general case where interaction among the interfaces is possible, we may therefore write:
\begin{equation}\label{domegacstv}
 \du \Omega = - S \du T - N \du\mu + 2\gamma(L_z)\du A
\end{equation} 

Note that  $\Omega$ is  a first order homogeneous function of the interface area (cf. \Eq{domega}). This
allows us to write:
\begin{equation}\label{omegaextensive}
   \Omega = 2\gamma A - \Pper A L_z
\end{equation} 
This equation will be exploited later on  in order to calculate $\gamma$.

In order to make contact with the standard thermodynamics of slit pores,
two new definitions must be introduced.  Firstly, a solvation force is defined as the
excess of $\Pper$ over the expected bulk pressure  $f_s(L_z)=\Pper-p$. 
Second, we introduce a surface free energy per unit area, $2\omega(L_z)=-(\Ppar-p)L_z$. With these definitions,
\Eq{domega} then becomes:
\begin{equation}\label{domegaslit}
\du \Omega = - S \du T -p \du V - N \du\mu + 2\omega\du A - A f_s \du L_z
\end{equation} 
The resemblance with the thermodynamics of slit pores is now apparent, although 
the volume here is just that of the actual system, $V=L_xL_yL_z$ with no mention of 
the reservoir. The equivalence of the two descriptions follows  by considering
the surface excess free energy. The  free energy increment of a bulk system of equal shape and size
is readily obtained either from  \Eq{domega} after replacement of $N$, $S$ and $\Ppar$ and $\Pper$ with the
corresponding bulk values or from \Eq{domegaslit} by retaining the first three terms in the right hand side.
In either case, we obtain the excess free energy differential that is standard in slit pore 
thermodynamics \cite{evans87}. 
\begin{equation}
 \du \Delta\Omega_{ex} = -\Delta S_{ex}\du T - \Delta N_{ex} \du\mu + 2\omega\du A - A f_s \du L_z
\end{equation} 
where subscript $ex$ stands for excess amounts over the corresponding bulk property.

Whether one uses \Eq{domega} or \Eq{domegaslit} as the fundamental relation is just a matter of convenience.
In experimental work on slit pores, one has control over $L_z$, $A$ is difficult to modify and
\Eq{domegaslit} is  more convenient. In the case of the simulations performed in this work 
\Eq{domega} and \Eq{domegacstv} will prove more useful since
only the independent variables are made explicit and none of these refer to the reservoir, which
cannot be actually simulated. 
Furthermore, we expect $\gamma(L_z)$ to converges fast.
Indeed, to a good approximation the difference between $\gamma(L_z)$ and
$\gamma_{\infty}$ arises from contributions well inside the infinitely large slit were $\Ppar - \Pper$ is zero
anyway. 

\section{New method for the calculation of interfacial tensions}

\subsection{Formal aspects}

The method proposed is conceptually very simple, and resembles other known methods where the simulation's box
shape is allowed to fluctuate \cite{finn88,finn89,forsman97,noe07}. In the absence of any other constraint, the shape of the system
at fixed temperature, chemical potential and volume will result from a competition between $\Ppar$ and $\Pper$. 
The rules of this competition are dictated by  a probability density, $f_{\mu V T}(A)$, and the corresponding
free energy, as follows:
\begin{equation}\label{probansatz}
        \Omega_{\mu V T}(A) = - k_B T \ln f_{\mu V T}(A)
\end{equation} 
This provides the link between the statistical mechanical property, $f$ and the macroscopic observables. By
use of \Eq{omegaextensive}, we find right away:
\begin{equation}\label{thermodistribution}
      f_{\mu V T}(A) = e^{-\beta( 2\gamma A - \Pper V )}
\end{equation} 
This is the key result of the method proposed: sampling the interface at constant volume from the above
distribution will readily yield the interfacial tension.
For systems where the interfaces do not interact and both $\gamma$ and $\Pper$ are system size independent, 
it is clear that
the slope of $\ln f_{\mu V T}$ yields the interfacial tension right away. For finite systems, $f_{\mu V T}$
the analysis is complicated by the non-trivial dependence of $\gamma(L_z)$ and also $\Pper(L_z)$. If sufficiently
fine data ara available, however, one can still recover $\gamma(L_z)$, since, according to \Eq{domegacstv} and
\Eq{probansatz}, we have
\begin{equation}\label{derivative}
            \frac{1}{f}\derpar{f}{A}{\mu V T} = - 2\beta\gamma(L_z)
\end{equation} 

In practice, however, \Eq{thermodistribution} as stated above is hardly useful, because $f_{\mu V T}(A)$ is a monotonous
function; i.e., the unconstrained system does not have an equilibrium state and during a simulation
it will either stretch
infinitely in order to eliminate the interface if $\gamma$ is positive, or increase its lateral size
continuously if  $\gamma$ is positive.

In order to overcome this difficulty, two different solutions seem possible. One is to add an extra work 
contribution  that  counterbalances the interfacial tension (balance sampling), yielding a proper equilibrium 
state. The other one
is to constraint the system's area within a bracketed interval chosen conveniently  (bracketed sampling). In practice, both 
physical situations  amount to the introduction of a modified or constrained probability density of the form:
\begin{equation}
       \widetilde{f}_{\mu V T}(A) = e^{-\beta W(A)} f_{\mu V T}(A)
\end{equation} 
where $W(A)$ is a suitably chosen function of $A$.

This distribution may be obtained from knowledge of the molecular interactions by  sampling  a modified grand canonical probability density of the
form:
\begin{equation}\label{distribution}
     \widetilde{f}_{\mu V T}(N,A;\{\rvec{}{}\}_N) \propto e^{\beta\mu N} e^{-\beta(\mathcal{U}(\{\rvec{}{}\}_N) + W(A)) }
\end{equation} 
where $\mathcal{U}$ is the intermolecular potential energy, and $\{\rvec{}{}\}_N$ is the set of position
vectors that define the microstate of a system with overall $N$ molecules. An obvious choice for the evaluation of 
the above
probability density is the standard Grand Canonical Monte Carlo Method, with the addition of an extra box
shape sampling. The box shape sampling is performed in the same spirit as in the standard NpT simulations.
Each position vector $\rvec{i}{}=(r_x,r_y,r_z)$ is transformed into a dimensionless position vector
$\tvec{i}{}=(r_x/L_x,r_y/L_y,r_z/L_z)$ and the relevant density distribution to sample the shape of the
simulation box becomes:
\begin{equation}
     \widetilde{f}_{\mu V T}(N,A;\{\tvec{}{}\}_N) \propto e^{\beta\mu N} V^N e^{
     -\beta(\mathcal{U}(\{\tvec{}{}\}_N;L_x,L_y,L_z) + W(A)) }
\end{equation} 
where we have  made explicit the parametric dependence  of the intermolecular energy on the box shape
that results from the transformation.

\subsection{Practical implementation}

In this section we discuss the implementation of the proposed method using Monte Carlo simulations for the
sampling of \Eq{distribution}. Other possible choices such as the Wang--Landau sampling \cite{wang01b}, and Molecular Dynamics 
will not be discussed here but are also possible.

A simulation box of suitable size is chosen with initial values of $L_z=D$ about twice those of $L_x=L_y=L$.
The simulation is then carried out using a standard grand canonical method, organized in cycles of
several regular canonical and grand canonical trials. After the end of each cycle, a box deformation is
attempted as follows. First, an area increment $\Delta$ is chosen uniformly from the interval $[-\Delta_{max}
,\Delta_{max}]$ and a new trial area $A_n=A_o+\Delta$ is proposed. The new box parameters
are set accordingly as $L_n=A_n^{1/2}$ and $D_n=V/A_n$. The attempted move is then accepted with probability:
\begin{equation}
  P_{accept} = \min \left ( 1,e^{-\beta ( \mathcal{U}_n - \mathcal{U}_o + W_n - W_o)} \right)  
\end{equation} 
Obviously, the choice of  $W$ is a crucial issue for the performance of the method. At first sight, a work
contribution that competes against the $\gamma A$ term so as to produce a well defined equilibrium state seems
the most appropriate. Unfortunately, due to the small systems considered the equilibrium value of $A$ must
fall within a fairly limited interval. Whereas this is can be achieved in principle, in
practice it amounts to a good {\em a priori} knowledge of $\gamma$. Therefore, after several attempts with
possible $W$ functions, it was found that the most convenient procedure is simply to bracket the  interface
area within an interval chosen beforehand, by means of the following $W$ function:
\begin{equation}\label{bracket}
  e^{-\beta W(A)} =
  \left \{
   \begin{array}{ccc}
      0 && A < A_{min} \\
	1 && A_{min} < A < A_{max} \\
	0 &&  A > A_{max} \\
   \end{array}	
   \right .
\end{equation} 
Within the chosen interval, the method produces the probability distribution of \Eq{thermodistribution}. For
non--interacting interfaces, a logarithmic plot of $f_{\mu V T}(A)$ yields the interfacial tension right away, 
while for interacting interfaces, \Eq{derivative} would have to be employed in order to determine $\gamma(L_z)$.
In the former case, routine calculation of interfacial tensions may become somewhat cumbersome, and a simpler
data analysis might be desirable. One such possibility is to exploit the expectation value of the surface
area, which reads:
\begin{equation}\label{avgamma}
 \frac{\prom{A}{} - A_{1/2}}{\Delta A} = \frac{1}{2\beta\gamma\Delta A} -  \frac{1}{2}\frac{1+
 e^{-2\beta\gamma\Delta A} }{1 - e^{-2\beta\gamma\Delta A} } 
\end{equation} 
where $\Delta A = A_{max} -  A_{min}$ and $A_{1/2}=1/2 ( A_{max} +  A_{min})$. 
The root of the above equation is the maximum likelihood estimate of
$\gamma$ given $\prom{A}{}$ and may be solved by means of a
Newton-Raphson method. The first order solution,  $2\beta\gamma\Delta A=12(A_{1/2}-\prom{A}{})/\Delta A$, may be used
as an excellent first guess, producing convergence in less than four iterations. In fact, this approximation
is good in most practical situations, as it produces deviations that are less than 3\% from the
exact solution for values of $(A_{1/2}-\prom{A}{})/\Delta A$ in the range $[-0.1,0.1]$.  

So far, the methodology described applies for the calculation of interfacial tensions in an open system. The
choice of variables, $\mu$, $V$ and $T$ is appropriate for calculating tensions of a fluid against a wall.
Interfacial tensions between two coexisting phases are also of great importance, but the thermodynamic
description employed here is not applicable right away. An initial state with two such phases will spontaneously 
transform into one or
the other phase and completely eliminate the interface. This problem is solved immediately simply by replacing 
$\mu$ with $N$ as the natural variable, with the number of particles fixed within the immiscibility gap.
All the framework that was employed in this section is then immediately applicable by simply
replacing the grand free energy by the Helmholtz free energy, and ignoring the sampling over the 
particle number.

\section{Model and Results}

\subsection{Molecular models studied}

The bracketed interfacial area sampling described in the previous section will be applied to the study of four
different systems. Firstly, we will consider an  argon like model, with pair interactions that only
depend on the mutual distance, $r$ between the atoms and obey the usual Lennard--Jones potential:
\begin{equation}
V_{LJ}(r) = 4 \epsilon \left\{ \left(\frac{\sigma}{r}\right)^{12} -
                                \left(\frac{\sigma}{r}\right)^{6} \right\}
\end{equation}
Secondly, a  central force potential of the square well type will also be considered:
\begin{equation}
V_{SW}(r)= \left\{ \begin{array}{ll}
 \infty & r < \sigma  \\
  -\epsilon & \sigma \leq r < \lambda \sigma \\
0 & r \geq \lambda \sigma \\
\end{array}   \right.
\end{equation}
where $\lambda$ defines the width of the square well and $\sigma$ and $\epsilon$ are the usual range and
energy parameters of the models.

Thirdly, we will consider a well known bead--spring polymer model which has been extensively studied. In this
model, all beads interact with each other, whether in the same or different polymer by means of a
 truncated and shifted  Lennard--Jones potential with $R_c=2 \cdot 2^{1/6}\sigma$. Additionally, adjacent
beads within the same polymer are held together by means of a Finite Extensible non Elastic (FENE) potential
of the form:
\begin{equation}
\Phi(r)  =  
 - k_s R_{\infty} ^2 \ln ( 1 - \frac{r^2}{R_{\infty}^2}  )  
\end{equation}
where $k_s=15 \epsilon / \sigma^2$ plays the role of spring constant and $R_{\infty}=1.5\sigma$ is the maximum 
allowed displacement between bonded beads.  Whereas for the simple atomic fluids only the vapor--liquid
interfacial tensions will be calculated, for the polymer we will also report wall--fluid interface tensions.
The polymers will thus be confined between two parallel plates in the usual slit pore geometry, with
fluid--substrate interactions  of the form:
\begin{equation}\label{wall}
  V_{wall}(z) = A \left \{ \left(\frac{\sigma}{z}\right)^9 -
                                   \left(\frac{\sigma}{z}\right)^3
                   \right \}
\end{equation}
where $A$ plays the role of a Hamaker constant.

The last model considered in this work will be a chain of freely jointed  tangent hard sphere chains. In this case,
beads interact with each other via a hard sphere potential of the form 
\begin{equation}
V_{HS}(r)= \left\{ \begin{array}{ll}
 \infty & r < \sigma  \\
0 & r \geq  \sigma \\
\end{array}   \right.
\end{equation}
Monomer within the same chain have a fixed bond length of $\sigma$, and are thus bonded tangentially. The
only other constraint is the non--local hard sphere interaction between beads more than one segment apart, 
so that the chain is otherwise fully flexible. We will consider systems with slit pore geometry and a purely
repulsive and athermal wall of the form:
\begin{equation}\label{athermalwall}
  V_{wall}(z) = 
  \left\{ \begin{array}{ll}
   \infty & z < \sigma/2  \\
   0 & z \geq  \sigma/2 \\
   \end{array}   \right.
\end{equation}


\section{}
\subsection{Liquid--Vapor surface tensions}

\subsubsection{Surface tension of the Lennard--Jones Fluid}

The methodology proposed was first explored and tested using the Lennard--Jones fluid, since several other
methods have already been tested for this system. The interactions were truncated at $R_c=2.5\sigma$, with no
further shift of the potential (this corresponds to the spherically truncated ST Lennard--Jones model of
Trokhymchuk and Alejandre \cite{trokhymchuk99}). The simulated boxes had an initial size of 13x13x45,
containing a total of 2137 particles. The production stage was organized in cycles. Each such cycle consisted of 
2000 attempted canonical attempts to displace a particle. Half of these attempts were
standard metropolis displacements set to yield about 50\% acceptance. The 
remaining half consist of  biased deletion--insertion movements \cite{macdowell01}. First, a particle chosen at random
was deleted. The particle was then inserted in one among several randomly chosen
positions with a probability given by the corresponding Rosenbluth weights. The attempted displacement was
accepted or rejected following the configurational bias rules \cite{siepmann92}.
It is expected that this trial move  will accelerate the proper equilibration
between the vapor and liquid phases and provide a better sampling of the interface than the metropolis
movements alone.  
After the end of each cycle, a box deformation attempt was performed, with $\Delta_{att}$ set during the
equilibration period to yield about 50\% acceptance. Averages were collected over about half million cycles
to one million cycles.

\begin{figure}
\includegraphics[width=0.6\textwidth]{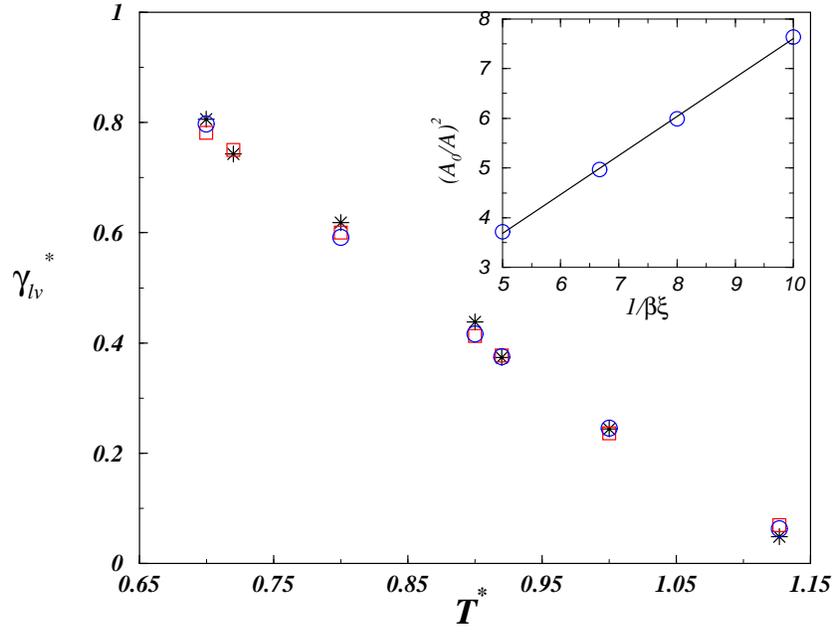}
\caption{\label{gammaLJST} Liquid--Vapor surface tension of the spherically truncated ($R_c=2.5$) Lennard
Jones model  $\gamma^*=\gamma_{lv}\sigma^2/\epsilon$ as a function of temperature $T^*=k_BT/\epsilon$.
The  squares and  circles are obtained  from the contertension and the bracketed sampling
methods, respectively. The stars are results from \protect\cite{trokhymchuk99}. The inset is a plot of inverse
squared surface area against $1/\beta\xi $, with slope equal to $\beta\gamma$.}
\end{figure}

Before considering the performance of the method proposed in the previous section, with the specific
bracketing choice for $W$ (cf. \Eq{bracket}), it is interesting to consider the performance of an alternative 
counter tension function selected so as to produce a proper equilibrium state. The surface tension
will favor states such that $\gamma A$ is minimum. In order to equilibrate this contribution, an extra term
acting in the opposite direction is needed. A possible choice that was considered is:
\begin{equation}
      W(A) = 2 \xi A_0^{2} / A
\end{equation} 
where $A_0 = V^{2/3}$ is introduced in the equation so that the parameter $\xi$ has the same dimensions as
$\gamma$. At constant $N$, $V$ and $T$, the relevant  Helmholtz free energy for the system in this case is simply:
\begin{equation}\label{mincon}
  F(A) = 2 \xi A_0^{2} / A + 2 \gamma A
\end{equation} 
Minimizing the free energy it is found that the equilibrium state of the surface area is such that:
\begin{equation}
  \frac{A}{A_0} = \left ( \frac{\xi}{\gamma} \right )^{1/2}
\end{equation} 
Therefore, the average interfacial area obtained from a simulation allows to calculate the surface tension.
An obvious way to test the above equation is by performing several simulations with different $\xi$ values. A
plot of $(A/A_0)^2$ against $\xi$ will produce a straight line with slope $\gamma$. This test is shown in
the inset of Fig.\ref{gammaLJST} for the ST--LJ model at $T=0.85\epsilon/k_B$. The plot clearly shows a straight line with 
slope
$\beta\gamma=(0.779\pm 0.010)\sigma^2$, in good agreement with the $\beta\gamma=(0.772\pm 0.010)\sigma^2 $ result suggested by Trokhymchuk and Alejandre from explicit
evaluation of the virial \cite{trokhymchuk99}. 
In that figure, results obtained by the balance method for several other temperatures is also shown. Good
agreement with Ref.\onlinecite{trokhymchuk99} is found in most cases. A more detailed analysis of the results,
together with the average surface area and $\xi$ values employed is reported in Table \ref{tabgammaLJ}. 
This is an important parameter, because $\gamma$ is known to show a significant dependence on the lateral
system size \cite{chen95,orea05}. Our results were performed for  similar system sizes as those reported in
Ref.\onlinecite{trokhymchuk99}  and the comparison shows good agreement. 
The only large discrepancy is found for the relatively high temperature $T=1.127\epsilon/k_B$, close to the
critical point. A reliable estimate of $\gamma$ at such high temperature is difficult for several reasons. The
interface width diverges close to the critical point, the bulk densities are attained only asymptotically slow
and the results will depend also on the perpendicular length $L_z$ and the length between the two interfaces
as well. One advantage of the method proposed here is that the interface is allowed to fluctuate, so that the
value of $\gamma$ that is obtained is averaged out over a finite interval of lateral lengths. For high
temperatures these fluctuations can be large. Indeed, the interfacial areas sampled during the simulation of
the  system at $T=1.127\epsilon/k_B$ for $\xi=0.0075\epsilon/\sigma^2$ fall broadly in an interval of lateral
lengths between 10 and 14$\sigma$. Unfortunately, these large fluctuations could actually be more a matter of 
concern than an advantage. For a very large system, the fluctuations about the equilibrium interface area 
will be
Gaussian, and the average interfacial area will be equal to the extremum of the distribution. For finite
systems as in our simulations, the large fluctuations imply that the interface area is actually sampling a non
Gaussian skewed distribution of the form $f(x)\sim\exp(-(ax + b/x))$, with $x$ the random variable and $a$ and $b$ some
coefficients. In such a case, the average and the extremum may deviate significantly, and
\Eq{mincon}  might not be a reliable estimate of the surface tension.

However, the most important problem with the counter tension methodology is in fact the choice of $\xi$.
Note that the average surface area that resulted from our simulations is such that the lateral length remains
close to the initial value $L=13\sigma$ for most temperatures (cf. Table \ref{tabgammaLJ}). Obviously, a look at 
\Eq{mincon} shows that this cannot be achieved accidentally. For an arbitrary choice of $\xi$, the resulting
equilibrium geometry may produce problems. For any $\xi$ larger than $\gamma$, it is clear that the average
equilibrium surface is larger than $A_0$. As a result, the lateral length becomes larger than the
perpendicular length, and the system becomes metastable: i.e., the liquid slab will have tendency to rotate by
$\pi/2$ so as to achieve a minimal surface area. For values of $\xi$ that are too large, on the contrary, the
simulation box will stretch very much and produce a long slab with very small surface area. The surface
tension that is obtained will then strongly depend on the equilibrium state, while use of \Eq{mincon} which
was derived by ignoring the system size dependence of $\gamma$ could be again not adequate. Obviously, one can
always choose $\xi$ such that $\prom{A}{}$ falls within some desired range as it was done in this work, but
this implies a reasonable first guess of $\gamma$, which is not always at hand.  

The bracketed interfacial area sampling proposed in the previous section overcomes all of these problems.
Firstly,  the sampling interval may be set at will right away, avoiding visits to undesired regions. Secondly,
no prior knowledge of $\gamma$ is required. Table \ref{tabgammaLJ} shows the results obtained using the
bracketing sampling strategy. Overall good agreement with both the virial and the balance sampling method is
found, though the results for the near to critical temperature $T=1.127\epsilon/k_B$ are again in conflict
with those obtained from the virial method. On the other hand, both methods employed in this work provide
results in fair agreement for that temperature, given that the lateral system sizes studied differ
considerably and the counter tension method could suffer from the problems discussed above. 

One important issue with the bracket sampling is the choice of bracketing interval $[A_{min},A_{max}]$. 
On the one hand, a large
interval is difficult to explore, because the states with large surface area are exponentially suppressed 
and so will suffer from poor sampling. On
the other hand,  large $\Delta y$ and $\Delta x$ values will lead to a small relative
error of the slope, $\Delta y/\Delta x$. Clearly, choosing the optimal value is not a trivial matter, although
we expect the optimal choice will depend on the dimensionless parameter $\xi=2\beta\gamma\Delta A$.
For $T=0.70\epsilon/k_B$ and  $T=1.00\epsilon/k_B$ two set of  simulations with different values of
$\xi$  were performed. In set $S$, a small interval was chosen, corresponding to a small
expected $\xi$ of about 1. On the contrary, in set $L$ we chose $\xi\sim 9$  such that
$f_{NVT}(A)$ varied about 3 orders of magnitude. The statistical analysis based on 
partial results of $2\beta\gamma$ obtained from \Eq{avgamma} after averaging the surface over 500,000 cycles remain
inconclusive. The error as measured from the standard deviation of the sample is very similar in both sets,
and no significant differences in the mean are observed either. For the state point at $T=0.70\epsilon/k_B$
set S yields $2\beta\gamma=(2.245\pm 0.050)\sigma^{-2}$ while two independent runs of set L produced 
$2\beta\gamma=(2.282\pm 0.040)\sigma^{-2}$ and $2\beta\gamma=(2.277\pm 0.027)\sigma^{-2}$. Similarly, for
$T=1.00\epsilon/k_B$ set S produced $2\beta\gamma=(0.490\pm 0.015)\sigma^{-2}$ while set L yield
$2\beta\gamma=(0.498\pm 0.017)\sigma^{-2}$ and $2\beta\gamma=(0.490\pm 0.011)\sigma^{-2}$.

\subsection{Surface tension of the Square--Well fluid}

Calculation of the surface tension from the anisotropy of the pressure tensor is particularly difficult for
fluids with discontinuous potentials. In such cases, there appears a delta Dirac in the virial expression for
every discontinuity in the potential. In practice, this means that the instantaneous virial can only be
measured approximately and that the computer program must be tailored for each model. The method of interface
sampling has the advantage of not requiring explicit evaluation of the virial, and this may prove particularly
advantageous in this case.

In this section the simple square--well fluid with $\lambda=1.5\sigma$ is chosen as a  prototypical
discontinuous model potential. Simulations were performed on systems containing overall 2222 atoms and initial
system size of 13x13x45. The sampling strategy is exactly the same as with the Lennard--Jones fluid, with
cycles of 2222 center of mass and configurational bias moves in the ratio 50:50.

The difficulties in sampling the square well fluid interface are illustrated by considering the
counter tension method explained in the previous section. In order to achieve a reasonable acceptance rate for
the box deformations, the maximal random displacement has to be set to very small values, and 
it becomes difficult for the system to reach a meaningful equilibrium value of the
interface area. For a temperature of $T=1.00\epsilon/k_B$ a value of $\Delta=0.035\sigma^2$ was 
required to
achieve the prescribed 50\% acceptance ratio in box deformations. On the other hand, a value of $\Delta=0.65
\sigma^2$ was
obtained for the Lennard--Jones fluid at the lowest temperature studied. Assuming a simple random walk for the
interface displacements, this would imply that the square well fluid needs $(0.65/0.035)^2$  more cycles than the
Lennard--Jones in order to sample the same surface area. In this conditions, the counter tension method becomes
inefficient and the bracket sampling strategy with a  small interval of about $0.5\sigma^2$ is more
convenient.  Even so, the simulations were performed over
ten million cycles in order to achieve  error bars comparable to those obtained for the Lennard--Jones.

The results obtained using the bracket sampling are shown in table \ref{tabSW}. Good agreement with previous
results using the virial method,  the Test Area Method  and Binder's method is
found \cite{orea03,singh03,gloor05}. 

In the Test Area method, a simulation
is carried out in an NVT ensemble of fixed shape. Every cycle, a `test' deformation at constant volume is
performed, and the corresponding change in configurational energy, $\Delta U$ is measured. Following Gloor et
al., it can be shown that the surface tension is given as:
\begin{equation}\label{tam}
\gamma_{lv} = \lim_{\Delta A\to 0} - k_B T \frac{\ln\prom{e^{-\beta \Delta U}}{0}}{\Delta A}
\end{equation} 
where the subscript "0" denotes that the interface is sampled  over the unperturbed system.
In the limit where the method becomes exact, it becomes equivalent to measuring the anisotropy of the
pressure tensor, with the virial evaluated numerically instead of analytically. Indeed, for small enough
deformations, the instantaneous energy change may be written as:
\begin{equation}
 \Delta U(\{\rvec{}{N}\}) = \derpar{U(\{\rvec{}{N}\})}{A}{V} \Delta A
\end{equation} 
Accordingly, an asymptotically exact Taylor expansion of Eq.(\ref{tam}) yields right away
\begin{equation}
   \gamma_{lv} = \prom{\derpar{U(\{\rvec{}{N}\})}{A}{V}}{0}
\end{equation} 
From the canonical partition function one can show that the analytical derivative in this equation is just the
anisotropy of the pressure tensor.
Both the TAM method and the interface sampling proposed in this work have the advantage of avoiding the
explicit calculation of the virial. However, the interface sampling method could avoid difficulties
encountered with the TAM method due to the nonequivalence between compressing and expanding the interface in
discontinuous systems \cite{gloor05}. For  very small deformations 
the three methods should become equivalent, although the virial method would seem to yield results
with smaller error bars \cite{demiguel06}.

\subsection{Fluid--substrate interfacial tension of Lennard--Jones chains}

In this section the interfacial tensions of the bead--spring polymer model described at the begining of this
section are considered. The interfacial properties of chains with 10 beads  at a subcritical temperature of
$T=1.68\epsilon/k_B$ have been studied previously and are well
known \cite{mueller00,varnik00}. The liquid--vapor surface tension has been calculated using Binder's method, while the spreading
coefficients $\gamma_{wv} - \gamma_{wl}$ were also evaluated with a related method all the way from the
wetting to the drying transition \cite{mueller00}. The interfacial tensions of the glassy polymer in narrow slit pores have 
also been measured using the virial method \cite{varnik00}. 

Simulations for the polymer model were carried out in the grand canonical ensemble. The chemical
potential was fixed to its coexistence value and the initial box size was  set to 13.8x13.8x27.6 (i.e., for
the liquid state, this amounts to about 315 polymers in the simulation box).  The configurational space was
sampled by means of center of mass displacements, standard canonical configurational bias displacements and
grand canonical configurational bias  insertion/deletion attempts in the ratio 25:25:50. For the vapor phase
only the last two kinds of trial moves  were attempted. The maximum allowed box
deformation attempts were adjusted during the equilibration period so as to provide roughly a 
50\% acceptance ratio. Averages were collected over roughly 1.5 million cycles for wall--liquid tensions and
about 5 million cycles for the wall--vapor tensions.

The results obtained for both the wall--liquid and wall--vapor interface tensions may be found in Table
\ref{tabFENELJ}.
Also included are the differences $\gamma_{wv} - \gamma_{wl}$ previously calculated using two different
methods. The comparison shows excellent agreement for the spreading coefficient and thus provides strong
support for the methodology. The liquid--vapor surface tension was also calculated using the interface
sampling in the canonical ensemble with a 13.5x13.5x50 box geometry and 185 molecules. The result,
$\gamma_{lv}=(0.156\pm0.008)\epsilon/\sigma^2$ is in good agreement with previous measurements using Binder's
method, $\gamma_{lv}=0.160\epsilon/\sigma^2$ \cite{mueller00}. 

For many practical applications, the relevant property is not really the surface
tensions alone, but the corresponding spreading coefficient, which governs both the wetting behavior of a
single interface and the capillary condensation/evaporation phenomenology. Figure \ref{gammaFENE} shows the results
of the spreading coefficient as a function of $\epsilon_w$. The points were the spreading coefficient cross
the liquid--vapor surface tensions indicate the location of wetting/drying transitions.  Independent measurement 
of the
tensions remains interesting, because it allows to compare the free energy of the wall--fluid interface
relative to that of a bulk system of equal volume. Such data is useful for comparison with self--consistent
and density functional theories, but can be also relevant in process were the interface is free to grow, i.e.,
nucleation or self assembly of biological systems. Table \ref{tabFENELJ} shows that both $\gamma_{wv}$ and
$\gamma_{wl}$ are negative for strongly attractive walls but change sign as the effective Hamaker constant
decreases. For positive but small values of $\epsilon_w$ the weak   attractive interactions
are not able to compensate for the free energy cost of a inhomogeneous density profile. The fact that the
liquid phase requires a wall strength larger than $\epsilon_w=2\epsilon$ before it becomes negative is consistent with  previously observed
density profiles, which  remain considerably depleted for $\epsilon_w$ as large
as $3\epsilon$ and only start to develop monomer like oscillatory behavior at about that value. Actually,
$\gamma_{wl}$ is about one order of  magnitude larger than $\gamma_{wv}$ all the way from $\epsilon_w=0$ to
$\epsilon_w=3.30$, and dominates completely the behavior of $\gamma_{wv}-\gamma_{wl}$. At $\epsilon_w=0$ both
the wall--vapor and wall--liquid interfaces are unfavorable relative to the corresponding bulk phases, and the
drying transition is actually driven by the very unfavorable wall--liquid interface. 

\subsection{Interfacial tension of chains of tangent hard spheres at an athermal wall}

Having studied the performance of the method for fluids with discontinuous potentials (square well) and for
polymer chains (FENE--LJ), we now consider a fluid made of tangent hard sphere chains adsorbed on purely
repulsive athermal walls. This system has been subject of great interest in the last decade, as a prototypical
model for  the study of polymer--wall interfaces, as well as a test bead for density functional theories
\cite{dickman88,yethiraj95b,hooper00}. 

We have  considered chains of  2, 4, 8, 12 and 16 monomers adsorbed on a hard wall. For each chain, we have
calculated the interface tensions corresponding to six different bulk densities of approximately
0.1, 0.2, 0.3, 0.4, 0.5 and 0.6  monomers per $\sigma^3$. We first performed bulk NVT simulations for those
densities, and evaluated the chemical potential by means of Widom's test particle method. For the largest chains
at heigh density, Widom's test failed and the grand potential could not be properly evaluated. An arbitrary
value was imposed in those cases, based on the approximation that the excess chemical potential is linear in
the chain length.  The values of the chemical potentials  employed in the simulations are collected in Table
\ref{chempot}.
Once the desired chemical potential was known, grand canonical simulations were performed on a system with initial 
size of 15x15x35. The grand canonical distribution function was sampled by using center of mass, rotation,
configurational bias and grand canonical configurational bias insertion/deletion moves in the ratio
15:15:35:35, except for dimers, were the ratio was set to 15:15:0:70. After suitable equilibration, averages
were collected over 3 to 7 million cycles, with longer runs required for the highest densities.

The choice of bracketing interval in this case is also a matter of concern, since the conditions change over a
wide interval, and the trial displacements required to achieve 50\% acceptance in the box shape moves varies
from about $3\sigma^2$ to about $3\times 10^{-2} \sigma^2$ as the monomer density increases from
$0.1\sigma^{-3}$ to $0.6\sigma^{-3}$ (this parameter being much less sensitive to chain length). At this stage
we developed a systematic method to decide the size of the lateral size sampling interval which works well.
A short simulation of about 100000 cycles is performed, and  the lateral size mean squared displacement is
calculated. A plot as a function of Monte Carlo cycles shows an initial linear and steady growth corresponding
to the unconstrained diffusion process,  followed by a plateau occasioned by the reflecting boundary
conditions on the lateral size (Fig. \ref{msd}). 
Measuring an effective diffusion coefficient, $D_{eff}$, from the slope of the mean
squared displacement within the first regime allows to estimate the expected lateral size displacement within
a given number of cycles. Obviously, the lateral size needs to have reflected
several many times if meaningful averages are to be collected. Therefore, the size of the sampling interval
required for the condition to be obeyed within a prescribed amount, $n_{\rm block}$ of MC cycles may be 
estimated from $D_{eff}$. In practice, we chose the sampling interval from the distance traveled 
during 10000 cycles.

The results of our calculations are presented in Table \ref{tabTHSC}. As seen from the table, the surface
tensions could be evaluated with relative errors lying between 1 and 10\%, with accuracy deteriorating at the
highest densities. Inspection of the results as plotted in
Fig.\ref{gammaTHSC} shows the interface tensions are always
positive, increase with monomer density and decrease with chain length. Simulation results suggest an
asymptotic behavior for long chains at low densities. For higher densities, however, the error bars are too
large and this trend cannot be confirmed. 

Hooper et al. have calculated  interface tensions of athermal chains with no intramolecular excluded volume
interactions \cite{hooper00}. These authors find that the interface tensions of a 20-mer first increases with density, but then gradually decreases and actually becomes negative at heigh densities.
The different behavior is related to the choice of dividing surface (cf. Eq.(\ref{athermalwall}). In this work
the dividing surface coincides with the actuall wall surface (i.e., $z=0$), while Hooper et al. assume that
the dividing surface is located at the distance of the closest approach of a segment to the surface (i.e.,
$z=\sigma/2$). 
Our results can be easily recalculated to another choice of the dividing surface
by using the simple relation  $\gamma_{z=\sigma/2} = \gamma_{z=0} - \frac{1}{2}p\sigma$,
where the subscripts stand for the location of the
dividing surface and $p$ is the bulk pressure (cf. Eq.~(4.3) in Ref.~\onlinecite{hooper00}).
This equation shows that our choice of dividing surface yields what is sometimes known
as the intermolecular contribution to the interfacial tension. For the proposed method, the choice of dividing
surface is implied in the volume definition.\cite{demiguel06} 
Here we have employed $V = A\times L_z$, while restricting the
system's  volume to that space available to the molecule's center of mass we would require $V=A\times
(L_z-\sigma)$ (a slit pore is used in practice, so $\frac{1}{2}\sigma$ must be substracted twice). 
These definitions dictate the length to breadth ratio of attempted deformations in the Monte
Carlo simulation, and thus produce interfacial tensions according to the choice of dividing surface. In order
to illustrate this point, we have performed some simulations for chains of length $m=12$ and a dividing
surface located at $z=\sigma/2$.  Fig.\ref{newdividingsurfacem12} presents a plot of the simulation results,
clearly showing the non--monotomic trend observed by Hooper et al \cite{hooper00}. Included as squared symbols
are results for $\gamma_{z=\sigma/2}$ obtained from our previously calculated $\gamma_{z=0}$ and shifted by 
$-\frac{1}{2}p\sigma$. The results agree within the error interval.


Along with the simulation results, we also include predictions from three different theoretical treatments.
The first approach is based on  Scaled Particle Theory \cite{reiss59}, which has been shown to be accurate for 
fluids of
spherical symmetry, whether hard spheres \cite{heni99,demiguel06}, or Lennard--Jones \cite{bresme02}. 
In this approach one considers  the free
energy change that results from inserting a large particle, and identifies the chemical potential with
macroscopic contributions arising from $pV$ and surface work \cite{reiss60,henderson83,bresme02}:
\begin{equation}
\du\mu_{\sigma} = 2\pi\gamma\sigma\du\sigma + \frac{\pi}{2}p\sigma^2\du\sigma
\end{equation} 
where $\mu_{\sigma}$ denotes the chemical potential of the large particle of diameter $\sigma$ 
inserted in the fluid. The macroscopic interface tension for a flat interface is obtained from the above 
equation in the limit of
infinite dilution and particle diameter as a term proportional to $\sigma$. Here we consider a fluid mixture
of tangent hard sphere chains and large hard sphere particles as described by Wertheim's perturbation theory\cite{wertheim87}, 
together with the Boublik
equation of state for the hard sphere reference mixture \cite{boublik70}.
After some tedious algebra, we obtain:
\begin{equation}\label{spt:1}
\pi\beta\sigma^2\gamma =  \frac{1}{(1+\eta)^2(1-\eta)^2}\left\{ 
\begin{array}{c}
(3-6\eta^2+3\eta^4)\ln(1-\eta) + 3\eta + 14\eta^2-5\eta^4-\eta^5+\eta^6 \\
+  (3\eta-5\eta^2+2\eta^4+\eta^5-\eta^6)/m 
\end{array}
\right \}
\end{equation}

The second approach is based on microscopic density functional theory for polymeric fluids \cite{woodward90,kierlik93,yu02}.
Within this framework the grand potential of the system is a functional of the local density of polymer
$\rho({\bf R})$, 
\begin{eqnarray}\label{dft:1}
\Omega[\rho({\bf R})]&=&\int d{\bf R} \rho({\bf R})(V_{ext}({\bf R})-\mu)+
\beta^{-1}\int  d{\bf R} \rho({\bf R})
(\ln(\rho({\bf R}))-1)\nonumber\\
&&+\int  d{\bf R} V_{b}({\bf R})\rho({\bf R})+F_{ex}[\rho_{PS}({\bf r})]\;.
\end{eqnarray}
In the above ${\bf R}=({\bf r}_1,{\bf r}_2,\ldots,{\bf r}_m)$ denotes the set of segment positions,
$V_{ext}$ is the external potential, while $V_b$ is the bonding potential that satisfies
$\exp [-\beta V_{b}({\bf R})]=
\prod_{i=1}^{m-1} (\delta(|{\bf r}_{i+1}-{\bf r}_{i}|-\sigma))/(4\pi\sigma^{2})$.
$F_{ex}[\rho_{PS}({\bf r})]$
is the excess free energy approximated as a functional of the average segment density defined
as $\rho_{PS}({\bf r})=\sum_{i=1}^{m}\int\!\!d{\bf R}\delta({\bf r}-{\bf r}_i)\rho({\bf R})$.
In this work we apply the Yu and Wu functional for $F_{ex}$, which is based
on fundamental measures' theory of Rosenfeld \cite{rosenfeld89,yu02}. According to this approach
the excess free energy is a volume integral $\beta F_{ex}=\int d{\bf r}\Phi$. 
The excess free energy density $\Phi$ is a simple function of a set of
weighted densities $\{n_{\alpha}\}$, $\alpha=3,2,1,0,V1,V2$ defined as
\begin{equation}
n_{\alpha}({\bf r})=\int\!\!d{\bf r}' \rho_{PS}({\bf r}')
w_{\alpha}({\bf r}-{\bf r}')\;.
\end{equation}
The weight functions $w_{\alpha}$ depend on geometrical properties of the segments
and are given explicitely in Refs.~\onlinecite{rosenfeld89,yu02}.

In the Yu and Wu theory the excess free energy density
is represented as a sum of two contributions, $\Phi=\Phi_{HS}+\Phi_P$.
$\Phi_{HS}$ describes the reference mixture of hard spheres and
is based on the Boublik
equation of state\cite{boublik70,Roth02,yu02:2}
\begin{eqnarray}\label{dft:2}
\Phi_{HS}=-n_0 \ln (1-n_{3})+
\frac{n_{1}n_{2}-{\bm n}_{V1}\cdot 
{\bm n}_{V2}}{1-n_{3}}&&\nonumber\\
+(n_2^3-3n_2{\bm n}_{V2}\cdot{\bm n}_{V2})\frac{n_{3}+
(1-n_3)^2\ln (1-n_3)} 
{36\pi (n_3)^2(1-n_{3})^{2}}&&\,,
\end{eqnarray}
while the excess free energy density due to the chain connectivity $\Phi_{P}$ is 
an ``inhomogeneous counterpart'' of the perturbation term in TPT1
\begin{equation} \label{dft:3}
\Phi_{P}=\frac{1-m}{m}n_0\zeta\ln[y_{HS}]\;,
\end{equation}
where $\zeta=1-\mathbf{n}_{V2} \cdot \mathbf{n}_{V2}/(n_2)^2$.
$y_{HS}$ is the expression
for the contact value of the hard-sphere radial distribution function
\begin{equation}\label{dft:4}  
y_{HS}=\frac 1{1-n_3}+\frac{n_2\sigma\zeta }{%
4(1-n_3)^2}+\frac{(n_2\sigma)^2\zeta }{72(1-n_3)^3}\;.
\end{equation}

The functional outlined above can be minimized numerically using the variational principle
$\delta \Omega /\delta \rho({\bf R})$ = 0. The resulting equilibrium density
profile can be inserted into \Eq{dft:1} yielding the grand potential which can be used
to calculate the interfacial tension.

The third theoretical approach utilizes the bulk limit of the microscopic density functional theory
outlined above. Starting point is the same homogenous one-component polymeric fluid as in the SPT
route, and again we consider the change of the grand potential, $\beta\Delta\Omega$, 
due to the insertion of a single big hard sphere into the system.
This change can be expressed in terms of the derivatives of the excess free energy density
\begin{equation}
\beta\Delta\Omega= 
\frac{\partial \Phi}{\partial n_{3}}~\zeta_3+
\frac{\partial \Phi}{\partial n_{2}}~\zeta_2+
\frac{\partial \Phi}{\partial n_{1}}~\zeta_1+
\frac{\partial \Phi}{\partial n_{0}}~\zeta_0 \;,
\end{equation} 
where $\zeta_i$ $i=3,2,1,0$ are characteristic functions of the shape of the inserted big particle \cite{bryk03,Koenig04,Roth06}.
In particular $\zeta_3$ and $\zeta_2$ are respectively the volume and the area of the big inserted hard sphere. If the functional was fully consistent with the SPT theory, the volume-dependent term $\partial \Phi/\partial n_{3}$ should give the pressure, 
while the area-dependent term $\partial \Phi/\partial n_{2}$ should give the planar wall-fluid
interfacial tension ($\partial \Phi/\partial n_{1}$ and $\partial \Phi/\partial n_{0}$ should
give the curvature-dependent corrections to the surface tension\cite{bryk03,Koenig04,Roth06}).
The functional given by Eqs.~(\ref{dft:2}-\ref{dft:4}) is not fully consistent with SPT but accepting this 
deficiency we can still identify the area-dependent term with the planar wall-fluid surface tension.
Noting that in the bulk limit the vector weighted densities ${\bf n}_{V2}$ and ${\bf n}_{V1}$ vanish
and $n_3\to\eta$, $n_2\to 6\eta/\sigma$, $n_1\to 3\eta/{\sigma}^2\pi$, $n_0\to 6\eta/{\sigma}^3\pi$ the interfacial
tension for tangent hard-sphere-polymer fluid at a hard wall is given as
\begin{equation}\label{eqpawel}
\pi\beta\sigma^2\gamma= 
\pi\sigma^2\frac{\partial \Phi}{\partial n_{2}}=\frac{3\eta(2-\eta)}{(1-\eta)^2}+
3\ln(1-\eta)+\frac{1-m}{m}\frac{2\eta(3-\eta)}{(4-2\eta)}\;.
\end{equation}
The first two terms in the r.h.s of Eq.\ref{eqpawel} represent the interface tension for the
hard-sphere fluid at a hard wall and is the same as in the SPT approach presented above.
The last contribution arises due to the chain connectivity but it differs from that of the
SPT approach.  
Figure 5 presents the results for the interfacial tension of tangent hard sphere chains
at a hard wall as a function of the chain length $m$. The sets of curves
were calculated for the bulk segment densities $\rho_\sigma^3=0.1$, 0.2, 0.3,
0.4, 0.5 and 0.6 (from bottom to top, respectively).
The circles denote the interface sampling simulation data. The solid lines are the 
results obtained by numerical minimization of the DFT, while  the dashed and dotted
lines denote, respectively, the SPT approach (\Eq{spt:1}) and the DFT-based analytical approximation
(\Eq{eqpawel}). We note that the DFT results are in overall good agreement with
simulations. The SPT approach always underestimates the interface tension and the agreement
is only qualitative. The DFT-based analytical approximation gives poor results
at low density but it significantly improves at higher density. 
Note that, as mentioned earlier, the interface tension is always positive because the
dividing surface is set at the actual wall surface. This choice is particularly useful for the theoretical
treatment, since adding the $-\frac{1}{2}p\sigma$ extra contribution consistent with the choice of dividing surface
at $z=\frac{1}{2}\sigma$ would obviously yield a less tractable equation. Figure \ref{newdividingsurfacem12}
shows results for the interface tensions that result from this definition for chains of length $m=12$.
Consistent with our previous observations, all three theoretical treatments provide qualitative agreement with
simulations, though the Yu and Wu DFT theory and the corresponding DFT-based analytical approximation are
in better agreement than the SPT approach.

\section{Conclusions}

We have developed a  method for the efficient calculation of interface tensions by means of 
computer simulations. The method allows the interfacial area to wander randomly and extracts the interface
tension from the resulting probability distribution.
This strategy shares common features with Binder's method and perturbative
methods such as those proposed by Bresme and Quirke and Gloor et al \cite{binder82,bresme99,gloor05}.  
On the one hand it relies on histogram analysis or the related average of a sampled probability distribution;
on the other hand the sampling is the result of small finite deformations of the simulation box.
The method avoids the explicit calculation of the pressure tensor and is therefore of 
great generality. It can be applied likewise for simple or complex fluids with continuous or discontinuous 
potentials, and may be employed for interface tension calculations of free or bound interfaces. 
Any sampling method for the gand canonical or canonical distributions is appropriate, and 
improved sampling over wide ranges of box shape could be implemented using known techniques such as
multicanonical \cite{berg92},
successive sampling \cite{virnau04b}, transition matrix \cite{fitzgerald99},  or Wang--Landau
sampling \cite{wang01b}.

The proposed method was tested for the Lennard--Jones and Square Well fluids and good agreement was obtained
with results found in the literature for similar system sizes. 
For systems with continuous potential, our results show that the method yields similar error bars at the same
computational cost. The method runs into problems when the surface area  cannot be sampled efficiently
over a large interval. This is the case for the Square Well fluid, were somewhat greater error bars were found
compared to the virial method at similar computational cost. On the other hand, the method is advantageous for
the calculation of wall--substrate interfacial tensions at low density. In such cases, a large surface area
can be sampled easily, and  low error bars result. In such cases, calculation of the interface
tension via the pressure tensor
anisotropy is a problem,  because the small number of collision events prevents accurate estimation of the
virial.

The technique was applied to the calculation of wall--fluid interfacial tensions for a  bead--spring Lennard
Jones chain, providing results for both the wall--vapor and wall--liquid tensions. The related spreading
coefficient was found to be in good agreement with previous results \cite{mueller00}. We also studied a system
of tangent hard--sphere chains adsorbed on athermal walls for a wide range of chain lengths and densities. 
At constant monomer density the interface tension approaches an asymptotic constant value from above. For
constant chain length, the interface tension increases regularly with density. The results where used to test
three different theories. Density Functional Theory as proposed by Yu and Wu \cite{yu02}, an analytical approximation
based on that theory and another analytic result inspired on Scaled Particle Theory \cite{reiss59}. 
The latter approach
reproduces the known accuracy for monomers, but  deteriorates significantly with increasing chain length. On
the other hand, the first two approaches provide rather good agreement. Particularly, the analytical result of
Eq.(\ref{eqpawel}) is surprisingly simple and yields predictions of similar good quality for all chain lengths
studied.
 

\begin{acknowledgments}
LGM wishes to thank  Marcus M\"uller, F. J. Blas, E. de Miguel, F. Bresme and C. Vega for helpful discussions;
This work was supported by Ministerio de Educacion y Ciencia through a Ramon y Cajal grant and project 
FIS2007–66079-C02-00 and by Comunidad Autonoma de Madrid throught project MOSSNOHO-S0505/ESP/0299.
PB acknowledges EU for partial funding this work as a TOK contract No. 509249.
\end{acknowledgments}


\begin{thebibliography}{70}
\expandafter\ifx\csname natexlab\endcsname\relax\def\natexlab#1{#1}\fi
\expandafter\ifx\csname bibnamefont\endcsname\relax
  \def\bibnamefont#1{#1}\fi
\expandafter\ifx\csname bibfnamefont\endcsname\relax
  \def\bibfnamefont#1{#1}\fi
\expandafter\ifx\csname citenamefont\endcsname\relax
  \def\citenamefont#1{#1}\fi
\expandafter\ifx\csname url\endcsname\relax
  \def\url#1{\texttt{#1}}\fi
\expandafter\ifx\csname urlprefix\endcsname\relax\def\urlprefix{URL }\fi
\providecommand{\bibinfo}[2]{#2}
\providecommand{\eprint}[2][]{\url{#2}}

\bibitem[{\citenamefont{Rowlinson and Widom}(1982)}]{rowlinson82b}
\bibinfo{author}{\bibfnamefont{J.}~\bibnamefont{Rowlinson}} \bibnamefont{and}
  \bibinfo{author}{\bibfnamefont{B.}~\bibnamefont{Widom}},
  \emph{\bibinfo{title}{Molecular Theory of Capillarity}}
  (\bibinfo{publisher}{Clarendon}, \bibinfo{address}{Oxford},
  \bibinfo{year}{1982}).

\bibitem[{\citenamefont{Seemann et~al.}(2005)\citenamefont{Seemann, Brinkmann,
  Kramer, Lange, and Lipowsky}}]{seemann05}
\bibinfo{author}{\bibfnamefont{R.}~\bibnamefont{Seemann}},
  \bibinfo{author}{\bibfnamefont{M.}~\bibnamefont{Brinkmann}},
  \bibinfo{author}{\bibfnamefont{E.~J.} \bibnamefont{Kramer}},
  \bibinfo{author}{\bibfnamefont{F.~F.} \bibnamefont{Lange}}, \bibnamefont{and}
  \bibinfo{author}{\bibfnamefont{R.}~\bibnamefont{Lipowsky}},
  \bibinfo{journal}{Proc. Nat. Acad. Sci.} \textbf{\bibinfo{volume}{102}},
  \bibinfo{pages}{1848} (\bibinfo{year}{2005}).

\bibitem[{\citenamefont{Milchev et~al.}(2005)\citenamefont{Milchev, Muller, and
  Binder}}]{milchev05}
\bibinfo{author}{\bibfnamefont{A.}~\bibnamefont{Milchev}},
  \bibinfo{author}{\bibfnamefont{M.}~\bibnamefont{Muller}}, \bibnamefont{and}
  \bibinfo{author}{\bibfnamefont{K.}~\bibnamefont{Binder}},
  \bibinfo{journal}{Phys. Rev. E} \textbf{\bibinfo{volume}{72}},
  \bibinfo{pages}{031603} (\bibinfo{year}{2005}).

\bibitem[{\citenamefont{MacDowell and M{\"u}ller}(2006)}]{macdowell06}
\bibinfo{author}{\bibfnamefont{L.~G.} \bibnamefont{MacDowell}}
  \bibnamefont{and}
  \bibinfo{author}{\bibfnamefont{M.}~\bibnamefont{M{\"u}ller}},
  \bibinfo{journal}{J. Chem. Phys.} \textbf{\bibinfo{volume}{124}},
  \bibinfo{pages}{084907} (\bibinfo{year}{2006}).

\bibitem[{\citenamefont{Rowlinson}(1979)}]{rowlinson79}
\bibinfo{author}{\bibfnamefont{J.~S.} \bibnamefont{Rowlinson}},
  \bibinfo{journal}{J. Stat. Phys.} \textbf{\bibinfo{volume}{20}},
  \bibinfo{pages}{197} (\bibinfo{year}{1979}).

\bibitem[{\citenamefont{Cahn and Hilliard}(1958)}]{cahn58}
\bibinfo{author}{\bibfnamefont{J.~W.} \bibnamefont{Cahn}} \bibnamefont{and}
  \bibinfo{author}{\bibfnamefont{J.~E.} \bibnamefont{Hilliard}},
  \bibinfo{journal}{J. Chem. Phys.} \textbf{\bibinfo{volume}{28}},
  \bibinfo{pages}{258} (\bibinfo{year}{1958}).

\bibitem[{\citenamefont{Tarazona}(1984)}]{tarazona84}
\bibinfo{author}{\bibfnamefont{P.}~\bibnamefont{Tarazona}},
  \bibinfo{journal}{Mol. Phys.} \textbf{\bibinfo{volume}{52}},
  \bibinfo{pages}{81} (\bibinfo{year}{1984}).

\bibitem[{\citenamefont{Woodward}(1990)}]{woodward90}
\bibinfo{author}{\bibfnamefont{C.~E.} \bibnamefont{Woodward}},
  \bibinfo{journal}{J. Chem. Phys.} \textbf{\bibinfo{volume}{94}},
  \bibinfo{pages}{3183} (\bibinfo{year}{1990}).

\bibitem[{\citenamefont{M{\"u}ller and MacDowell}(2000)}]{mueller00}
\bibinfo{author}{\bibfnamefont{M.}~\bibnamefont{M{\"u}ller}} \bibnamefont{and}
  \bibinfo{author}{\bibfnamefont{L.~G.} \bibnamefont{MacDowell}},
  \bibinfo{journal}{Macromolecules} \textbf{\bibinfo{volume}{33}},
  \bibinfo{pages}{3902} (\bibinfo{year}{2000}).

\bibitem[{\citenamefont{Bryk and Sokolowsky}(2004)}]{bryk04}
\bibinfo{author}{\bibfnamefont{P.}~\bibnamefont{Bryk}} \bibnamefont{and}
  \bibinfo{author}{\bibfnamefont{S.}~\bibnamefont{Sokolowsky}},
  \bibinfo{journal}{J. Chem. Phys.} \textbf{\bibinfo{volume}{121}},
  \bibinfo{pages}{11314} (\bibinfo{year}{2004}).

\bibitem[{\citenamefont{Bryk et~al.}(2004)\citenamefont{Bryk, Bucior,
  Sokolowski, and Zukocinski}}]{bryk04b}
\bibinfo{author}{\bibfnamefont{P.}~\bibnamefont{Bryk}},
  \bibinfo{author}{\bibfnamefont{K.}~\bibnamefont{Bucior}},
  \bibinfo{author}{\bibfnamefont{S.}~\bibnamefont{Sokolowski}},
  \bibnamefont{and}
  \bibinfo{author}{\bibfnamefont{G.}~\bibnamefont{Zukocinski}},
  \bibinfo{journal}{J. Physics-condensed Matter} \textbf{\bibinfo{volume}{16}},
  \bibinfo{pages}{8861} (\bibinfo{year}{2004}).

\bibitem[{\citenamefont{Irving and Kirkwood}(1950)}]{irving50}
\bibinfo{author}{\bibfnamefont{J.~H.} \bibnamefont{Irving}} \bibnamefont{and}
  \bibinfo{author}{\bibfnamefont{J.~G.} \bibnamefont{Kirkwood}},
  \bibinfo{journal}{J. Chem. Phys.} \textbf{\bibinfo{volume}{18}},
  \bibinfo{pages}{17} (\bibinfo{year}{1950}).

\bibitem[{\citenamefont{Magda et~al.}(1985)\citenamefont{Magda, Tirrell, and
  Davis}}]{magda85}
\bibinfo{author}{\bibfnamefont{J.~J.} \bibnamefont{Magda}},
  \bibinfo{author}{\bibfnamefont{M.}~\bibnamefont{Tirrell}}, \bibnamefont{and}
  \bibinfo{author}{\bibfnamefont{H.~T.} \bibnamefont{Davis}},
  \bibinfo{journal}{J. Chem. Phys.} \textbf{\bibinfo{volume}{83}},
  \bibinfo{pages}{1888} (\bibinfo{year}{1985}).

\bibitem[{\citenamefont{Trokhymchuk and Alejandre}(1999)}]{trokhymchuk99}
\bibinfo{author}{\bibfnamefont{A.}~\bibnamefont{Trokhymchuk}} \bibnamefont{and}
  \bibinfo{author}{\bibfnamefont{J.}~\bibnamefont{Alejandre}},
  \bibinfo{journal}{J. Chem. Phys.} \textbf{\bibinfo{volume}{111}},
  \bibinfo{pages}{8510} (\bibinfo{year}{1999}).

\bibitem[{\citenamefont{Orea et~al.}(2003)\citenamefont{Orea, Duda, and
  Alejandre}}]{orea03}
\bibinfo{author}{\bibfnamefont{P.}~\bibnamefont{Orea}},
  \bibinfo{author}{\bibfnamefont{Y.}~\bibnamefont{Duda}}, \bibnamefont{and}
  \bibinfo{author}{\bibfnamefont{J.}~\bibnamefont{Alejandre}},
  \bibinfo{journal}{J. Chem. Phys.} \textbf{\bibinfo{volume}{118}},
  \bibinfo{pages}{5635} (\bibinfo{year}{2003}).

\bibitem[{\citenamefont{del Rio and de~Miguel}(1997)}]{martin97}
\bibinfo{author}{\bibfnamefont{E.~M.} \bibnamefont{del Rio}} \bibnamefont{and}
  \bibinfo{author}{\bibfnamefont{E.}~\bibnamefont{de~Miguel}},
  \bibinfo{journal}{Phys. Rev. E} \textbf{\bibinfo{volume}{55}},
  \bibinfo{pages}{2916} (\bibinfo{year}{1997}).

\bibitem[{\citenamefont{Varnik et~al.}(2000)\citenamefont{Varnik, Baschnagel,
  and Binder}}]{varnik00}
\bibinfo{author}{\bibfnamefont{F.}~\bibnamefont{Varnik}},
  \bibinfo{author}{\bibfnamefont{J.}~\bibnamefont{Baschnagel}},
  \bibnamefont{and} \bibinfo{author}{\bibfnamefont{K.}~\bibnamefont{Binder}},
  \bibinfo{journal}{J. Chem. Phys.} \textbf{\bibinfo{volume}{113}},
  \bibinfo{pages}{4444} (\bibinfo{year}{2000}).

\bibitem[{\citenamefont{Milchev and Binder}(2001)}]{milchev01}
\bibinfo{author}{\bibfnamefont{A.}~\bibnamefont{Milchev}} \bibnamefont{and}
  \bibinfo{author}{\bibfnamefont{K.}~\bibnamefont{Binder}},
  \bibinfo{journal}{J. Chem. Phys.} \textbf{\bibinfo{volume}{114}},
  \bibinfo{pages}{8610} (\bibinfo{year}{2001}).

\bibitem[{\citenamefont{Duque et~al.}(2004)\citenamefont{Duque, Pamies, and
  Vega}}]{duque04}
\bibinfo{author}{\bibfnamefont{D.}~\bibnamefont{Duque}},
  \bibinfo{author}{\bibfnamefont{J.~C.} \bibnamefont{Pamies}},
  \bibnamefont{and} \bibinfo{author}{\bibfnamefont{L.~F.} \bibnamefont{Vega}},
  \bibinfo{journal}{J. Chem. Phys.} \textbf{\bibinfo{volume}{121}},
  \bibinfo{pages}{11395} (\bibinfo{year}{2004}).

\bibitem[{\citenamefont{Binder}(1982)}]{binder82}
\bibinfo{author}{\bibfnamefont{K.}~\bibnamefont{Binder}},
  \bibinfo{journal}{Phys. Rev. A} \textbf{\bibinfo{volume}{25}},
  \bibinfo{pages}{1699} (\bibinfo{year}{1982}).

\bibitem[{\citenamefont{Potoff and Panagiotopoulos}(2000)}]{potoff00}
\bibinfo{author}{\bibfnamefont{J.~J.} \bibnamefont{Potoff}} \bibnamefont{and}
  \bibinfo{author}{\bibfnamefont{A.~Z.} \bibnamefont{Panagiotopoulos}},
  \bibinfo{journal}{J. Chem. Phys.} \textbf{\bibinfo{volume}{112}},
  \bibinfo{pages}{6411} (\bibinfo{year}{2000}).

\bibitem[{\citenamefont{Errington}(2003)}]{errington03}
\bibinfo{author}{\bibfnamefont{J.~R.} \bibnamefont{Errington}},
  \bibinfo{journal}{Phys. Rev. E} \textbf{\bibinfo{volume}{67}},
  \bibinfo{pages}{012102} (\bibinfo{year}{2003}).

\bibitem[{\citenamefont{MacDowell}(2003)}]{macdowell03b}
\bibinfo{author}{\bibfnamefont{L.~G.} \bibnamefont{MacDowell}},
  \bibinfo{journal}{J. Chem. Phys.} \textbf{\bibinfo{volume}{119}},
  \bibinfo{pages}{453} (\bibinfo{year}{2003}).

\bibitem[{\citenamefont{Hooper et~al.}(2000{\natexlab{a}})\citenamefont{Hooper,
  McCoy, Curro, and van Swol}}]{hooper00}
\bibinfo{author}{\bibfnamefont{J.~B.} \bibnamefont{Hooper}},
  \bibinfo{author}{\bibfnamefont{J.~D.} \bibnamefont{McCoy}},
  \bibinfo{author}{\bibfnamefont{J.~G.} \bibnamefont{Curro}}, \bibnamefont{and}
  \bibinfo{author}{\bibfnamefont{F.}~\bibnamefont{van Swol}},
  \bibinfo{journal}{J. Chem. Phys.} \textbf{\bibinfo{volume}{113}},
  \bibinfo{pages}{2021} (\bibinfo{year}{2000}{\natexlab{a}}).

\bibitem[{\citenamefont{Furukawa and Binder}(1982)}]{furukawa82}
\bibinfo{author}{\bibfnamefont{H.}~\bibnamefont{Furukawa}} \bibnamefont{and}
  \bibinfo{author}{\bibfnamefont{K.}~\bibnamefont{Binder}},
  \bibinfo{journal}{Phys. Rev. A} \textbf{\bibinfo{volume}{26}},
  \bibinfo{pages}{556} (\bibinfo{year}{1982}).

\bibitem[{\citenamefont{MacDowell et~al.}(2004)\citenamefont{MacDowell, Virnau,
  M{\"u}ller, and Binder}}]{macdowell04}
\bibinfo{author}{\bibfnamefont{L.~G.} \bibnamefont{MacDowell}},
  \bibinfo{author}{\bibfnamefont{P.}~\bibnamefont{Virnau}},
  \bibinfo{author}{\bibfnamefont{M.}~\bibnamefont{M{\"u}ller}},
  \bibnamefont{and} \bibinfo{author}{\bibfnamefont{K.}~\bibnamefont{Binder}},
  \bibinfo{journal}{J. Chem. Phys.} \textbf{\bibinfo{volume}{120}},
  \bibinfo{pages}{5293} (\bibinfo{year}{2004}).

\bibitem[{\citenamefont{M{\"u}ller and Schick}(1996)}]{mueller96}
\bibinfo{author}{\bibfnamefont{M.}~\bibnamefont{M{\"u}ller}} \bibnamefont{and}
  \bibinfo{author}{\bibfnamefont{M.}~\bibnamefont{Schick}},
  \bibinfo{journal}{J. Chem. Phys.} \textbf{\bibinfo{volume}{105}},
  \bibinfo{pages}{8282} (\bibinfo{year}{1996}).

\bibitem[{\citenamefont{Chacon and Tarazona}(2005)}]{chacon05}
\bibinfo{author}{\bibfnamefont{E.}~\bibnamefont{Chacon}} \bibnamefont{and}
  \bibinfo{author}{\bibfnamefont{P.}~\bibnamefont{Tarazona}},
  \bibinfo{journal}{J. Phys.: Condens. Matter} \textbf{\bibinfo{volume}{17}},
  \bibinfo{pages}{S3493} (\bibinfo{year}{2005}).

\bibitem[{\citenamefont{Reiss et~al.}(1959)\citenamefont{Reiss, Frisch, and
  Lebowitz}}]{reiss59}
\bibinfo{author}{\bibfnamefont{H.}~\bibnamefont{Reiss}},
  \bibinfo{author}{\bibfnamefont{H.~L.} \bibnamefont{Frisch}},
  \bibnamefont{and} \bibinfo{author}{\bibfnamefont{J.~L.}
  \bibnamefont{Lebowitz}}, \bibinfo{journal}{J. Chem. Phys.}
  \textbf{\bibinfo{volume}{31}}, \bibinfo{pages}{369} (\bibinfo{year}{1959}).

\bibitem[{\citenamefont{Henderson}(1983)}]{henderson83}
\bibinfo{author}{\bibfnamefont{J.~R.} \bibnamefont{Henderson}},
  \bibinfo{journal}{Mol. Phys.} \textbf{\bibinfo{volume}{50}},
  \bibinfo{pages}{741} (\bibinfo{year}{1983}).

\bibitem[{\citenamefont{Bryk et~al.}(2003)\citenamefont{Bryk, Roth, Mecke, and
  Dietrich}}]{bryk03}
\bibinfo{author}{\bibfnamefont{P.}~\bibnamefont{Bryk}},
  \bibinfo{author}{\bibfnamefont{R.}~\bibnamefont{Roth}},
  \bibinfo{author}{\bibfnamefont{K.~R.} \bibnamefont{Mecke}}, \bibnamefont{and}
  \bibinfo{author}{\bibfnamefont{S.}~\bibnamefont{Dietrich}},
  \bibinfo{journal}{Phys. Rev. E} \textbf{\bibinfo{volume}{68}},
  \bibinfo{pages}{031602} (\bibinfo{year}{2003}).

\bibitem[{\citenamefont{Bresme and Quirke}(1998)}]{bresme98}
\bibinfo{author}{\bibfnamefont{F.}~\bibnamefont{Bresme}} \bibnamefont{and}
  \bibinfo{author}{\bibfnamefont{N.}~\bibnamefont{Quirke}},
  \bibinfo{journal}{Phys. Rev. Lett.} \textbf{\bibinfo{volume}{80}},
  \bibinfo{pages}{3791} (\bibinfo{year}{1998}).

\bibitem[{\citenamefont{Bresme and Quirke}(1999)}]{bresme99}
\bibinfo{author}{\bibfnamefont{F.}~\bibnamefont{Bresme}} \bibnamefont{and}
  \bibinfo{author}{\bibfnamefont{N.}~\bibnamefont{Quirke}},
  \bibinfo{journal}{J. Chem. Phys.} \textbf{\bibinfo{volume}{110}},
  \bibinfo{pages}{3536} (\bibinfo{year}{1999}).

\bibitem[{\citenamefont{Gloor et~al.}(2005)\citenamefont{Gloor, Jackson, Blas,
  and de~Miguel}}]{gloor05}
\bibinfo{author}{\bibfnamefont{G.~J.} \bibnamefont{Gloor}},
  \bibinfo{author}{\bibfnamefont{G.}~\bibnamefont{Jackson}},
  \bibinfo{author}{\bibfnamefont{F.~J.} \bibnamefont{Blas}}, \bibnamefont{and}
  \bibinfo{author}{\bibfnamefont{E.}~\bibnamefont{de~Miguel}},
  \bibinfo{journal}{J. Chem. Phys.} \textbf{\bibinfo{volume}{123}},
  \bibinfo{pages}{134703} (\bibinfo{year}{2005}).

\bibitem[{\citenamefont{de~Miguel and Jackson}(2006)}]{demiguel06}
\bibinfo{author}{\bibfnamefont{E.}~\bibnamefont{de~Miguel}} \bibnamefont{and}
  \bibinfo{author}{\bibfnamefont{G.}~\bibnamefont{Jackson}},
  \bibinfo{journal}{Mol. Phys.} \textbf{\bibinfo{volume}{104}},
  \bibinfo{pages}{3717} (\bibinfo{year}{2006}).

\bibitem[{\citenamefont{V{\"o}rtler and Smith}(2000)}]{vortler00}
\bibinfo{author}{\bibfnamefont{H.~L.} \bibnamefont{V{\"o}rtler}}
  \bibnamefont{and} \bibinfo{author}{\bibfnamefont{W.~R.} \bibnamefont{Smith}},
  \bibinfo{journal}{J. Chem. Phys.} \textbf{\bibinfo{volume}{112}},
  \bibinfo{pages}{5168} (\bibinfo{year}{2000}).

\bibitem[{\citenamefont{Widom}(1963)}]{widom63}
\bibinfo{author}{\bibfnamefont{B.}~\bibnamefont{Widom}}, \bibinfo{journal}{J.
  Chem. Phys.} \textbf{\bibinfo{volume}{39}}, \bibinfo{pages}{2808}
  (\bibinfo{year}{1963}).

\bibitem[{\citenamefont{Chandler et~al.}(1986)\citenamefont{Chandler, McCoy,
  and Singer}}]{Chandler86:1}
\bibinfo{author}{\bibfnamefont{D.}~\bibnamefont{Chandler}},
  \bibinfo{author}{\bibfnamefont{J.~D.} \bibnamefont{McCoy}}, \bibnamefont{and}
  \bibinfo{author}{\bibfnamefont{S.~J.} \bibnamefont{Singer}},
  \bibinfo{journal}{J. Chem. Phys.} \textbf{\bibinfo{volume}{85}},
  \bibinfo{pages}{5971} (\bibinfo{year}{1986}).

\bibitem[{\citenamefont{Helfand}(1972)}]{helfand72}
\bibinfo{author}{\bibfnamefont{E.}~\bibnamefont{Helfand}}, \bibinfo{journal}{J.
  Chem. Phys.} \textbf{\bibinfo{volume}{56}}, \bibinfo{pages}{3592}
  (\bibinfo{year}{1972}).

\bibitem[{\citenamefont{M{\"u}ller and MacDowell}(2003)}]{mueller03b}
\bibinfo{author}{\bibfnamefont{M.}~\bibnamefont{M{\"u}ller}} \bibnamefont{and}
  \bibinfo{author}{\bibfnamefont{L.~G.} \bibnamefont{MacDowell}},
  \bibinfo{journal}{J. Phys.: Condens. Matter} \textbf{\bibinfo{volume}{15}},
  \bibinfo{pages}{R609} (\bibinfo{year}{2003}).

\bibitem[{\citenamefont{Frischknecht et~al.}(2002)\citenamefont{Frischknecht,
  Weinhold, Salinger, Curro, Frink, and McCoy}}]{Frischknecht02:1}
\bibinfo{author}{\bibfnamefont{A.~L.} \bibnamefont{Frischknecht}},
  \bibinfo{author}{\bibfnamefont{J.~D.} \bibnamefont{Weinhold}},
  \bibinfo{author}{\bibfnamefont{A.~G.} \bibnamefont{Salinger}},
  \bibinfo{author}{\bibfnamefont{J.~G.} \bibnamefont{Curro}},
  \bibinfo{author}{\bibfnamefont{L.~J.~D.} \bibnamefont{Frink}},
  \bibnamefont{and} \bibinfo{author}{\bibfnamefont{J.~D.} \bibnamefont{McCoy}},
  \bibinfo{journal}{J. Chem. Phys.} \textbf{\bibinfo{volume}{117}},
  \bibinfo{pages}{10385} (\bibinfo{year}{2002}).

\bibitem[{\citenamefont{Hooper et~al.}(2000{\natexlab{b}})\citenamefont{Hooper,
  McCoy, and Curro}}]{hooper00:1}
\bibinfo{author}{\bibfnamefont{J.~B.} \bibnamefont{Hooper}},
  \bibinfo{author}{\bibfnamefont{J.~D.} \bibnamefont{McCoy}}, \bibnamefont{and}
  \bibinfo{author}{\bibfnamefont{J.~G.} \bibnamefont{Curro}},
  \bibinfo{journal}{J. Chem. Phys.} \textbf{\bibinfo{volume}{112}},
  \bibinfo{pages}{3090} (\bibinfo{year}{2000}{\natexlab{b}}).

\bibitem[{\citenamefont{Yu and Wu}(2002{\natexlab{a}})}]{yu02}
\bibinfo{author}{\bibfnamefont{Y.~X.} \bibnamefont{Yu}} \bibnamefont{and}
  \bibinfo{author}{\bibfnamefont{J.~Z.} \bibnamefont{Wu}}, \bibinfo{journal}{J.
  Chem. Phys.} \textbf{\bibinfo{volume}{117}}, \bibinfo{pages}{2368}
  (\bibinfo{year}{2002}{\natexlab{a}}).

\bibitem[{\citenamefont{Kierlik and Rosinberg}(1993)}]{kierlik93}
\bibinfo{author}{\bibfnamefont{E.}~\bibnamefont{Kierlik}} \bibnamefont{and}
  \bibinfo{author}{\bibfnamefont{M.~L.} \bibnamefont{Rosinberg}},
  \bibinfo{journal}{J. Chem. Phys.} \textbf{\bibinfo{volume}{99}},
  \bibinfo{pages}{3950} (\bibinfo{year}{1993}).

\bibitem[{\citenamefont{Wertheim}(1987)}]{wertheim87}
\bibinfo{author}{\bibfnamefont{M.~S.} \bibnamefont{Wertheim}},
  \bibinfo{journal}{J. Chem. Phys.} \textbf{\bibinfo{volume}{87}},
  \bibinfo{pages}{7323} (\bibinfo{year}{1987}).

\bibitem[{\citenamefont{Evans and Marconi}(1987)}]{evans87}
\bibinfo{author}{\bibfnamefont{R.}~\bibnamefont{Evans}} \bibnamefont{and}
  \bibinfo{author}{\bibfnamefont{U.~M.~B.} \bibnamefont{Marconi}},
  \bibinfo{journal}{J. Chem. Phys.} \textbf{\bibinfo{volume}{87}},
  \bibinfo{pages}{7138} (\bibinfo{year}{1987}).

\bibitem[{\citenamefont{Finn and Monson}(1989)}]{finn89}
\bibinfo{author}{\bibfnamefont{J.~E.} \bibnamefont{Finn}} \bibnamefont{and}
  \bibinfo{author}{\bibfnamefont{P.~A.} \bibnamefont{Monson}},
  \bibinfo{journal}{Phys. Rev. A} \textbf{\bibinfo{volume}{39}},
  \bibinfo{pages}{6402} (\bibinfo{year}{1989}).

\bibitem[{\citenamefont{Finn and Monson}(1988)}]{finn88}
\bibinfo{author}{\bibfnamefont{J.~E.} \bibnamefont{Finn}} \bibnamefont{and}
  \bibinfo{author}{\bibfnamefont{P.~A.} \bibnamefont{Monson}},
  \bibinfo{journal}{Mol. Phys.} \textbf{\bibinfo{volume}{65}},
  \bibinfo{pages}{1345} (\bibinfo{year}{1988}).

\bibitem[{\citenamefont{Forsman and Woodward}(1997)}]{forsman97}
\bibinfo{author}{\bibfnamefont{J.}~\bibnamefont{Forsman}} \bibnamefont{and}
  \bibinfo{author}{\bibfnamefont{C.~E.} \bibnamefont{Woodward}},
  \bibinfo{journal}{Mol. Phys.} \textbf{\bibinfo{volume}{90}},
  \bibinfo{pages}{637} (\bibinfo{year}{1997}).

\bibitem[{\citenamefont{Almarza et~al.}()\citenamefont{Almarza, de~Miguel, and
  Jackson}}]{noe07}
\bibinfo{author}{\bibfnamefont{N.~G.} \bibnamefont{Almarza}},
\bibinfo{author}{\bibfnamefont{E.}~\bibnamefont{de~Miguel}},
\bibnamefont{and} \bibinfo{author}{\bibfnamefont{G.}~\bibnamefont{Jackson}},
\bibinfo{note}{private communication}.

\bibitem[{\citenamefont{Wang and Landau}(2001)}]{wang01b}
\bibinfo{author}{\bibfnamefont{F.~G.} \bibnamefont{Wang}} \bibnamefont{and}
  \bibinfo{author}{\bibfnamefont{D.~P.} \bibnamefont{Landau}},
  \bibinfo{journal}{Phys. Rev. Lett.} \textbf{\bibinfo{volume}{86}},
  \bibinfo{pages}{2050} (\bibinfo{year}{2001}).

\bibitem[{\citenamefont{MacDowell et~al.}(2001)\citenamefont{MacDowell, Vega,
  and Sanz}}]{macdowell01}
\bibinfo{author}{\bibfnamefont{L.~G.} \bibnamefont{MacDowell}},
  \bibinfo{author}{\bibfnamefont{C.}~\bibnamefont{Vega}}, \bibnamefont{and}
  \bibinfo{author}{\bibfnamefont{E.}~\bibnamefont{Sanz}}, \bibinfo{journal}{J.
  Chem. Phys.} \textbf{\bibinfo{volume}{115}}, \bibinfo{pages}{6220}
  (\bibinfo{year}{2001}).

\bibitem[{\citenamefont{Siepmann and Frenkel}(1992)}]{siepmann92}
\bibinfo{author}{\bibfnamefont{J.~I.} \bibnamefont{Siepmann}} \bibnamefont{and}
  \bibinfo{author}{\bibfnamefont{D.}~\bibnamefont{Frenkel}},
  \bibinfo{journal}{Mol. Phys.} \textbf{\bibinfo{volume}{75}},
  \bibinfo{pages}{59} (\bibinfo{year}{1992}).

\bibitem[{\citenamefont{Chen}(1995)}]{chen95}
\bibinfo{author}{\bibfnamefont{L.-J.} \bibnamefont{Chen}}, \bibinfo{journal}{J.
  Chem. Phys.} \textbf{\bibinfo{volume}{103}}, \bibinfo{pages}{10214}
  (\bibinfo{year}{1995}).

\bibitem[{\citenamefont{Orea et~al.}(2005)\citenamefont{Orea, Duda, and
  Alejandre}}]{orea05}
\bibinfo{author}{\bibfnamefont{P.}~\bibnamefont{Orea}},
  \bibinfo{author}{\bibnamefont{Duda}}, \bibnamefont{and}
  \bibinfo{author}{\bibnamefont{Alejandre}}, \bibinfo{journal}{J. Chem. Phys.}
  \textbf{\bibinfo{volume}{123}}, \bibinfo{pages}{114702}
  (\bibinfo{year}{2005}).

\bibitem[{\citenamefont{Singh et~al.}(2003)\citenamefont{Singh, Kofke, and
  Errington}}]{singh03}
\bibinfo{author}{\bibfnamefont{J.~K.} \bibnamefont{Singh}},
  \bibinfo{author}{\bibfnamefont{D.~A.} \bibnamefont{Kofke}}, \bibnamefont{and}
  \bibinfo{author}{\bibfnamefont{J.~R.} \bibnamefont{Errington}},
  \bibinfo{journal}{J. Chem. Phys.} \textbf{\bibinfo{volume}{119}},
  \bibinfo{pages}{3405} (\bibinfo{year}{2003}).

\bibitem[{\citenamefont{Dickman and Hall}(1988)}]{dickman88}
\bibinfo{author}{\bibfnamefont{R.}~\bibnamefont{Dickman}} \bibnamefont{and}
  \bibinfo{author}{\bibfnamefont{C.~K.} \bibnamefont{Hall}},
  \bibinfo{journal}{J. Chem. Phys.} \textbf{\bibinfo{volume}{89}},
  \bibinfo{pages}{3168} (\bibinfo{year}{1988}).

\bibitem[{\citenamefont{Yethiraj and Woodward}(1995)}]{yethiraj95b}
\bibinfo{author}{\bibfnamefont{A.}~\bibnamefont{Yethiraj}} \bibnamefont{and}
  \bibinfo{author}{\bibfnamefont{C.~E.} \bibnamefont{Woodward}},
  \bibinfo{journal}{J. Chem. Phys.} \textbf{\bibinfo{volume}{102}},
  \bibinfo{pages}{5499} (\bibinfo{year}{1995}).

\bibitem[{\citenamefont{Heni and L{\"o}wen}(1999)}]{heni99}
\bibinfo{author}{\bibfnamefont{M.}~\bibnamefont{Heni}} \bibnamefont{and}
  \bibinfo{author}{\bibfnamefont{H.}~\bibnamefont{L{\"o}wen}},
  \bibinfo{journal}{Phys. Rev. E} \textbf{\bibinfo{volume}{60}},
  \bibinfo{pages}{7057} (\bibinfo{year}{1999}).

\bibitem[{\citenamefont{Bresme}(2002)}]{bresme02}
\bibinfo{author}{\bibfnamefont{F.}~\bibnamefont{Bresme}}, \bibinfo{journal}{J.
  Phys. Chem. B} \textbf{\bibinfo{volume}{106}}, \bibinfo{pages}{7852}
  (\bibinfo{year}{2002}).

\bibitem[{\citenamefont{Reiss et~al.}(1960)\citenamefont{Reiss, Helfand, and
  Lebowitz}}]{reiss60}
\bibinfo{author}{\bibfnamefont{H.}~\bibnamefont{Reiss}},
  \bibinfo{author}{\bibfnamefont{H.~L. F.~E.} \bibnamefont{Helfand}},
  \bibnamefont{and} \bibinfo{author}{\bibfnamefont{J.~L.}
  \bibnamefont{Lebowitz}}, \bibinfo{journal}{J. Chem. Phys.}
  \textbf{\bibinfo{volume}{32}}, \bibinfo{pages}{119} (\bibinfo{year}{1960}).

\bibitem[{\citenamefont{Boublik}(1970)}]{boublik70}
\bibinfo{author}{\bibfnamefont{T.}~\bibnamefont{Boublik}}, \bibinfo{journal}{J.
  Chem. Phys.} \textbf{\bibinfo{volume}{53}}, \bibinfo{pages}{471}
  (\bibinfo{year}{1970}).

\bibitem[{\citenamefont{Rosenfeld}(1989)}]{rosenfeld89}
\bibinfo{author}{\bibfnamefont{Y.}~\bibnamefont{Rosenfeld}},
  \bibinfo{journal}{Phys. Rev. Lett.} \textbf{\bibinfo{volume}{63}},
  \bibinfo{pages}{980} (\bibinfo{year}{1989}).

\bibitem[{\citenamefont{Roth et~al.}(2002)\citenamefont{Roth, Evans, Lang, and
  Kahl}}]{Roth02}
\bibinfo{author}{\bibfnamefont{R.}~\bibnamefont{Roth}},
  \bibinfo{author}{\bibfnamefont{R.}~\bibnamefont{Evans}},
  \bibinfo{author}{\bibfnamefont{A.}~\bibnamefont{Lang}}, \bibnamefont{and}
  \bibinfo{author}{\bibfnamefont{G.}~\bibnamefont{Kahl}}, \bibinfo{journal}{J.
  Phys.: Condens. Matter} \textbf{\bibinfo{volume}{14}}, \bibinfo{pages}{12063}
  (\bibinfo{year}{2002}).

\bibitem[{\citenamefont{Yu and Wu}(2002{\natexlab{b}})}]{yu02:2}
\bibinfo{author}{\bibfnamefont{Y.~X.} \bibnamefont{Yu}} \bibnamefont{and}
  \bibinfo{author}{\bibfnamefont{J.~Z.} \bibnamefont{Wu}}, \bibinfo{journal}{J.
  Chem. Phys.} \textbf{\bibinfo{volume}{117}}, \bibinfo{pages}{10156}
  (\bibinfo{year}{2002}{\natexlab{b}}).

\bibitem[{\citenamefont{K{\"o}nig et~al.}(2004)\citenamefont{K{\"o}nig, Roth,
  and Mecke}}]{Koenig04}
\bibinfo{author}{\bibfnamefont{P.~M.} \bibnamefont{K{\"o}nig}},
  \bibinfo{author}{\bibfnamefont{R.}~\bibnamefont{Roth}}, \bibnamefont{and}
  \bibinfo{author}{\bibfnamefont{K.~R.} \bibnamefont{Mecke}},
  \bibinfo{journal}{Phys. Rev. Lett.} \textbf{\bibinfo{volume}{93}},
  \bibinfo{pages}{160601} (\bibinfo{year}{2004}).

\bibitem[{\citenamefont{Hansen-Goos and Roth}(2006)}]{Roth06}
\bibinfo{author}{\bibfnamefont{H.}~\bibnamefont{Hansen-Goos}} \bibnamefont{and}
  \bibinfo{author}{\bibfnamefont{R.}~\bibnamefont{Roth}}, \bibinfo{journal}{J.
  Phys.: Condens. Matter} \textbf{\bibinfo{volume}{18}}, \bibinfo{pages}{8413}
  (\bibinfo{year}{2006}).

\bibitem[{\citenamefont{Berg and Neuhaus}(1992)}]{berg92}
\bibinfo{author}{\bibfnamefont{B.~A.} \bibnamefont{Berg}} \bibnamefont{and}
  \bibinfo{author}{\bibfnamefont{T.}~\bibnamefont{Neuhaus}},
  \bibinfo{journal}{Phys. Rev. Lett.} \textbf{\bibinfo{volume}{68}},
  \bibinfo{pages}{9} (\bibinfo{year}{1992}).

\bibitem[{\citenamefont{Virnau and M{\"u}ller}(2004)}]{virnau04b}
\bibinfo{author}{\bibfnamefont{P.}~\bibnamefont{Virnau}} \bibnamefont{and}
  \bibinfo{author}{\bibfnamefont{M.}~\bibnamefont{M{\"u}ller}},
  \bibinfo{journal}{J. Chem. Phys.} \textbf{\bibinfo{volume}{120}},
  \bibinfo{pages}{10925} (\bibinfo{year}{2004}).

\bibitem[{\citenamefont{Fitzgerald et~al.}(1999)\citenamefont{Fitzgerald,
  Picard, and Silver}}]{fitzgerald99}
\bibinfo{author}{\bibfnamefont{M.}~\bibnamefont{Fitzgerald}},
  \bibinfo{author}{\bibfnamefont{R.~R.} \bibnamefont{Picard}},
  \bibnamefont{and} \bibinfo{author}{\bibfnamefont{R.~N.}
  \bibnamefont{Silver}}, \bibinfo{journal}{Europhysics Lett.}
  \textbf{\bibinfo{volume}{46}}, \bibinfo{pages}{282} (\bibinfo{year}{1999}).

\bibitem[{\citenamefont{Smit}(1995)}]{smit95b}
\bibinfo{author}{\bibfnamefont{B.}~\bibnamefont{Smit}}, \bibinfo{journal}{Mol.
  Phys.} \textbf{\bibinfo{volume}{85}}, \bibinfo{pages}{153}
  (\bibinfo{year}{1995}).

\end{thebibliography}

\clearpage

\newpage

\begin{table}
\caption{\label{tabgammaLJ} 
Surface tension for the Spherically Truncated Lennard--Jones potential as determined by the wandering interface 
method with balance and bracket sampling. Results obtained from the virial method are also shown for
comparison \cite{trokhymchuk99}.
Range column includes $\beta k$ for the counter tension method or the surface range
sampled, for the bracket method. Size column indicates either the average lateral size for
the wandering interface methods or the fixed lateral size for the virial method. Averages for balance sampling
were collected over 400 Kcycles, except for those indicated with an asterisk, collected over twice as many.
Averages for the bracket sampling were collected over 1000 Kcycles.}
 \begin{ruledtabular}
 \begin{tabular}{ccccc}
  $k_BT/\epsilon$     & method & constrain & system size  &  $\sigma^2\gamma_{lv}/\epsilon$   \\ 
  \hline
0.70 & balance & 0.276      & 13.88 & 0.781(8) \\
0.70 & bracket  & [169, 181] & 13.19 & 0.815(2) \\ 
0.70 & bracket  & [179, 183]  & 13.40 & 0.797(10) \\ 
0.70 & virial  & -          & 13.41 & 0.806(16) \\
0.72 & balance & 0.175      & 12.59 & 0.75(2) \\
0.72 & bracket  & [146, 158] & 12.11 & 0.735(4) \\       
0.72 & virial  &  -         & 13.41 & 0.743(3) \\
0.80$^*$ & balance & 0.200    & 14.16  & 0.594(5) \\  
0.80$^*$ & balance & 0.150    & 13.17  & 0.598(6) \\  
0.80$^*$ & balance & 0.125    & 12.57  & 0.601(6)  \\   
0.80   & balance & 0.100    & 11.83  & 0.61(2)  \\    
0.80   & bracket  & [179, 183] &13.40  & 0.591(7) \\
0.80 & virial  & -          & 13.41  & 0.618(9)  \\
0.90$^*$ & balance & 0.0756   & 12.54  & 0.413(3) \\
0.90   & bracket  & [178, 184] &13.38  & 0.416(9) \\
0.90   & virial & -         & 13.41  & 0.438(4) \\
0.92 & balance & 0.0903     & 13.49  &  0.377(7) \\  
0.92 & bracket  &  [177, 185] & 13.35  &  0.375(8) \\
0.92 & virial  & -          & 13.41  & 0.374(6) \\
1.00 & balance & 0.0466     & 13.14  & 0.236(6) \\   
1.00 & bracket  &  [172, 190] & 13.19  & 0.246(5) \\        
1.00 & virial  & -          & 13.41  & 0.244(5) \\
1.127$^*$ & bracket & [133, 139] & 11.65 & 0.0637(8)  \\   
1.127 & bracket & [170, 190]   & 13.29 & 0.0633(2)  \\   
1.127$^*$ & balance & 0.0057   & 10.97  &  0.070(3)  \\   
1.127$^*$ & balance & 0.0075   & 11.76  &  0.069(4)  \\  
1.127 & virial  & -          & 13.41  & 0.049(6)   \\ 
 \end{tabular}
 \end{ruledtabular}
\end{table}

\begin{table}
 \caption{\label{tabSW} Surface tension for the square well fluid with $\lambda=1.5\sigma$. 
  Results from this work are compared with the virial method (Ref.\onlinecite{orea03}) and with
  the Test Area Method (Ref.\onlinecite{gloor05}).
  }
 \begin{ruledtabular}
 \begin{tabular}{ccccc}
  $k_BT/\epsilon$     & method & constrain & system size  &  $\sigma^2\gamma_{lv}/\epsilon$   \\ 
0.90 & bracket & [173.75, 174.25]   & 13.19  & 0.431(40) \\
0.90 & virial &      & 10 & 0.423(10) \\
0.90 & TAM    &      & 10.76 & 0.468(18) \\
0.95 & bracket & [170.75, 171.25]   & 13.08   & 0.351(36) \\
0.95 & virial &      & 10 & 0.349(9) \\
0.95 & TAM    &      & 10.88 & 0.336(20) \\
1.00 & bracket & [170.7, 171.0]     & 13.07  & 0.256(20) \\
1.00 & virial &      & 10  & 0.268(13) \\
1.00 & TAM    &      & 11.01    & 0.288(20) \\
1.05 & bracket & [170.15, 170.45]     & 13.05  & 0.166(21) \\
1.05 & virial &      & 10  & 0.202(11) \\
1.05 & TAM    &      & 11.18    & 0.205(14) \\
1.10 & bracket & [168.75, 169.25]     & 13.00 & 0.132(22) \\
1.10 & virial &      & 10  & 0.142(12) \\
1.10 & TAM    &      & 11.39    & 0.153(14) \\
 \end{tabular}
 \end{ruledtabular}
\end{table}

\begin{table}
 \caption{\label{tabFENELJ} Wall--fluid interface  tension for  a bead--spring polymer model (FENE--LJ) at
 T=1.68. Interface tensions are given in units of the Lennard--Jones energy parameter $\epsilon$.
  }
 \begin{ruledtabular}
 \begin{tabular}{ccc}
 $\epsilon_w$ & $\gamma_{wl}$ & $\gamma_{wv}$ \\
3.30 & -0.181(6)  & -0.0139(3) \\
3.20 & -0.164(12) & -0.0119(3) \\
3.10 & -0.136(15) & -0.0106(3)  \\
3.00 & -0.120(10) & -0.0095(6) \\
2.80 & -0.098(7)  & -0.0072(1) \\
2.00 &  0.032(8)  & -0.0015(1) \\
1.00 &  0.161(10) & 0.0011(2)  \\
0.00 &   -        & -0.00009(10)  \\
 \end{tabular}
 \end{ruledtabular}
\end{table}

\begin{table}
 \caption{\label{chempot} 
 Chemical potentials ($\mu/k_BT$)  of tangent hard sphere chains for several chain lengths, $m$ and monomer densities
 $\rho$. The results for m=12 and m=16 at $\rho=0.60\sigma^{-3}$ could not be measured from the NVT+test
 particle method and merely correspond to the chemical potential imposed during the simulations. Results for
 $\mu/k_BT$ are given as an excess over ideal chains. The entry for zero density is the difference between
 real and ideal chain contributions (see Ref.\onlinecite{smit95b} for a discussion on reference states
 appropriate to configurational bias grand canonical simulations).
  }
 \begin{ruledtabular}
 \begin{tabular}{cccccc}
 $\rho$/m &  2 & 4 &  8  &   12  &    16  \\
 0.00 &   0 & 0.6277 & 2.1085 & 3.6495 & 5.2118 \\
 0.10 &  -2.34814(4) & -2.12764(7) & -0.8409(1) & 0.7589(2) & 2.4841(2) \\
 0.20 &  -0.81907(5) & -0.1513(1) & 1.9751(2) & 4.3995(6) & 6.9482(8) \\
 0.30 &  0.67658(5) & 2.0178(3) & 5.4555(4) & 9.19(2) & 13.060(2) \\
 0.40 &  2.4000(2) & 4.7275(3) & 10.123(2) & 15.81(7) & 21.59(7) \\
 0.50 &  4.5358(6) & 8.296(2) & 16.5(1) & 25.19(3) & 34.(2) \\
 0.60 &  7.308(1) & 13.126(5) & 25.2(4) & 43. & 54. \\
 \end{tabular}
 \end{ruledtabular}
\end{table}

\begin{table}
 \caption{\label{tabTHSC} Wall--fluid interface  tension for  tangent hard sphere chains on
 athermal substrates for several chain lengths, $m$ and monomer densities, $\rho$ (cf. Tab.\ref{chempot} for
 chemical potentials). 
 The entries of the table are interface tensions $\gamma/k_BT$. Results are given in units of the
 hard sphere diameter.
  }
 \begin{ruledtabular}
 \begin{tabular}{cccccc}
 $\rho$  & 2 & 4 &  8  &   12  &    16  \\
0.10   & 0.0447(5) & 0.0352(4) & 0.0295(4) & 0.0273(4) & 0.0248(2) \\
0.20   & 0.104(2) & 0.0905(10) & 0.0788(14) & 0.0755(10) & 0.0745(14) \\
0.30   & 0.197(6) & 0.159(6) & 0.148(4) & 0.149(5) & 0.153(3) \\
0.40  & 0.347(15) & 0.278(11) & 0.26(9) & 0.240(7) & 0.25(2) \\
0.50  & 0.44(3)  & 0.43(2)  & 0.40(3) & 0.375(11)  & 0.40(2)  \\
0.60  & 0.83(8)  & 0.63(3)  & 0.60(4) & 0.61(5) & 0.52(4)  \\
 \end{tabular}
 \end{ruledtabular}
\end{table}

\clearpage

\begin{figure}
\includegraphics[width=0.6\textwidth]{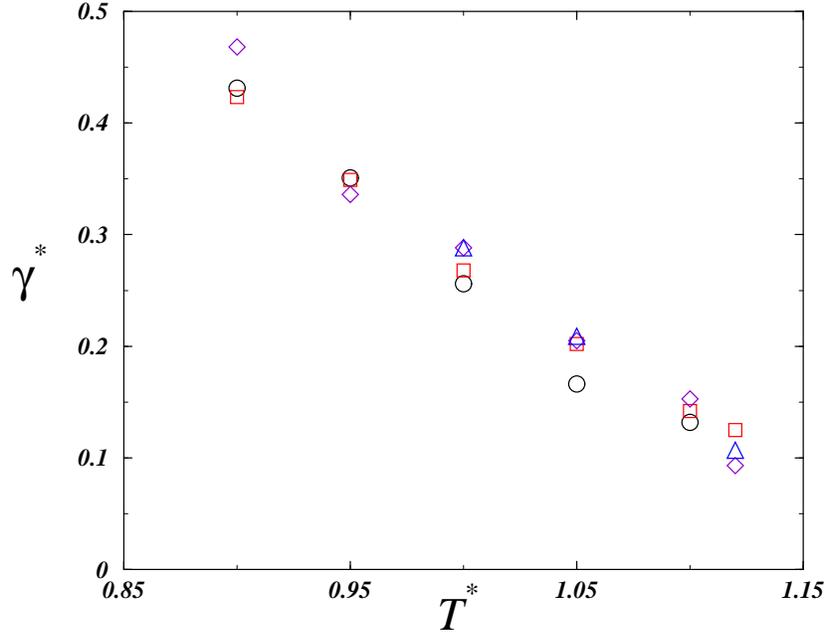}%
\caption{\label{gammaSWl1.5} Liquid--Vapor surface tension of a square well fluid with $\lambda=1.5\sigma$.
Results are given in reduced units,
with $\gamma^*=\gamma_{lv}\sigma^2/\epsilon$ plotted as a function of temperature $T^*=k_BT/\epsilon$.
Circles, results from the bracketed sampling method of this work. Squares, anisotropy of the virial
tensor \cite{orea03}. Diamonds, Test Area Method \cite{gloor05}. Triangles, Binder's method \cite{singh03}. 
}
\end{figure}

\begin{figure}
\includegraphics[width=0.6\textwidth]{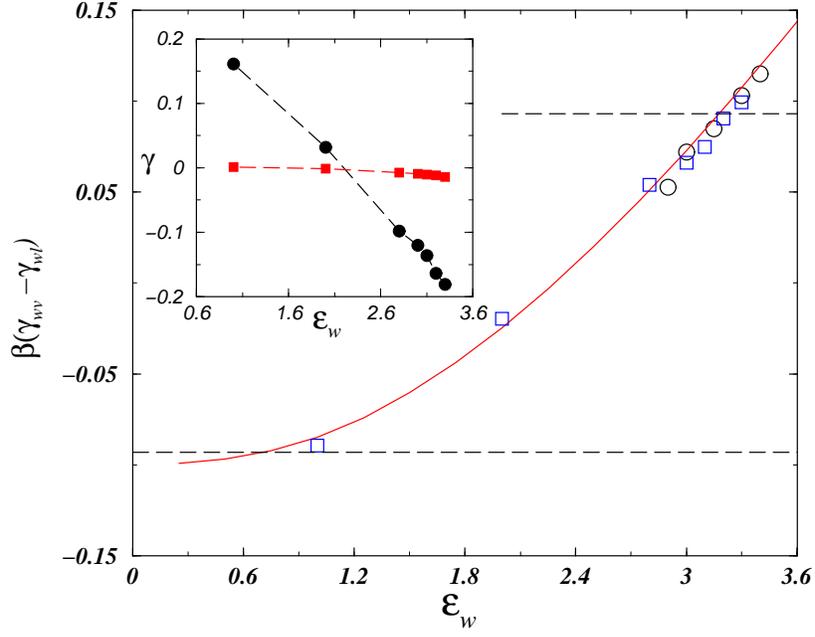}%
\caption{\label{gammaFENE}  Wall--fluid interface tensions for the LJ--FENE model. The figure shows the
spreading coefficient $\gamma_{wv}-\gamma_{wl}$  as a function of wall strength, $\epsilon_w$. Squares are
results obtained in this work from explicit evaluation of $\gamma_{wv}$ and $\gamma_{wl}$. Circles and lines
are results for the spreading coefficients from Ref.\onlinecite{mueller00} The dashed horizontal lines
correspond $\pm\gamma_{lv}$ and the intersections correspond to wetting and drying transitions.
The inset shows the tensions as a
function of wall strength. Circles and squares refer to the wall--liquid and wall--vapor tensions,
respectively.
}
\end{figure}

\begin{figure}
\includegraphics[width=0.6\textwidth]{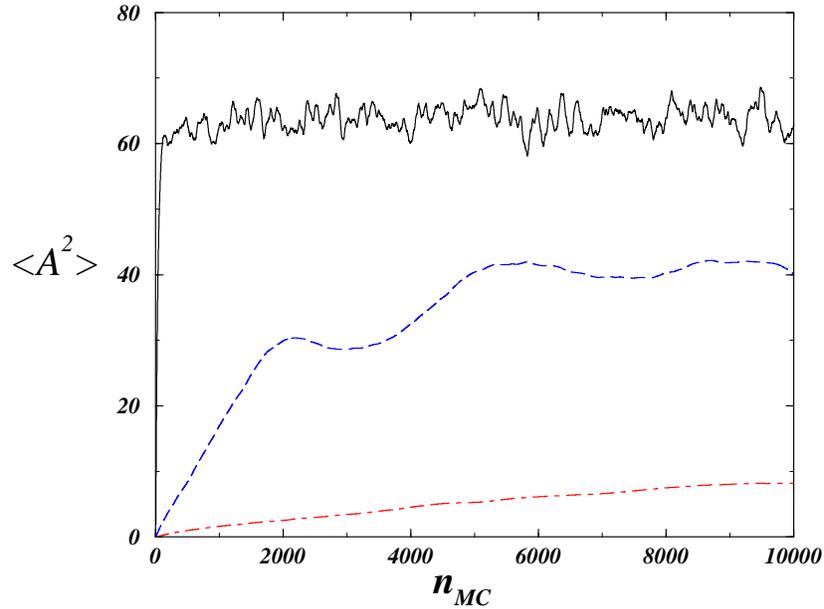}%
\caption{\label{msd}  Surface area mean squared displacement as a function of Monte Carlo cycles for chains of
length m=16. From top to bottom, results refer to monomer densities $\rho=0.10\sigma^{-3}$ (full line),
$\rho=0.30\sigma^{-3}$ (dashed line, 2 times magnified) and $\rho=0.50\sigma^{-3}$ (dash--dotted line, 10
times magnified).
}
\end{figure}

\begin{figure}
\includegraphics[width=0.6\textwidth]{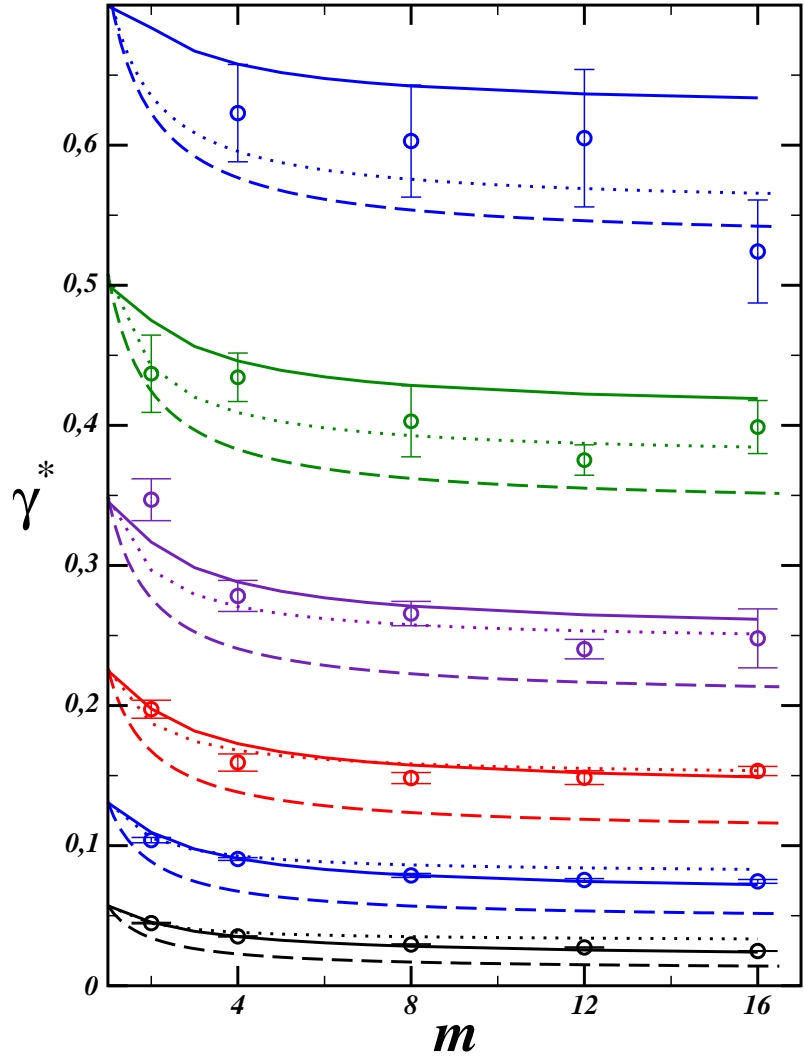}%
\caption{\label{gammaTHSC}  Wall--fluid interface tensions for hard sphere chains adsorbed on an athermal
substrate as a function of chain length. From top to bottom, results refer to $\rho=0.1\sigma^{-3}$,
$\rho=0.2\sigma^{-3}$, $\rho=0.3\sigma^{-3}$, $\rho=0.4\sigma^{-3}$, $\rho=0.5\sigma^{-3}$ and $\rho=0.6\sigma^{-3}$.
Symbols show simulation results from this work. Lines are theoretical results: Full line,
TPT1--DFT; dashed line, SPT; dotted lines, DFT-based analytical approximation.
}
\end{figure}

\begin{figure}
\includegraphics[width=0.6\textwidth,clip]{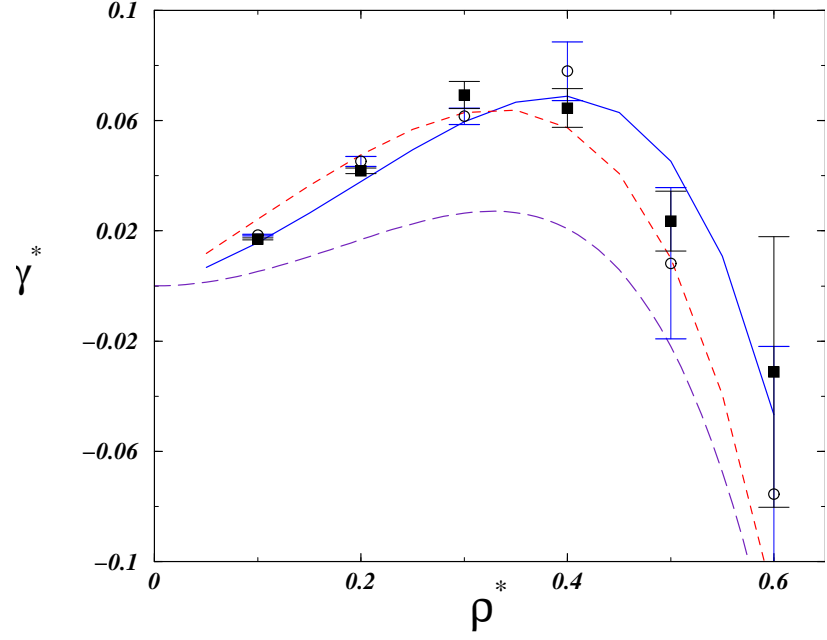}%
\caption{\label{newdividingsurfacem12}  Wall--fluid interface tensions for hard sphere chains of length $m=12$
adsorbed on an athermal
substrate with the dividing surface located at $z=\frac{1}{2}\sigma$. The circles are simulation results with
the volume of the slit pore defined as $V=(L_z-\sigma)$. The filled squares are results obtained by
substracting $\frac{1}{2}p\sigma$ to the simulated data for $\gamma_{z=0}$.
The lines are theoretical results.  Full line,
TPT1--DFT; dashed line, SPT; dotted lines, DFT-based analytical approximation.
}
\end{figure}

\end{document}